\documentstyle[12pt]{article}
\catcode`\@=11

\@addtoreset{equation}{section}
\def\theequation{\arabic{section}.\arabic{equation}}

\catcode`\@=11

\def\section{\@startsection{section}{1}{\z@}{3.5ex plus 1ex minus
   .2ex}{2.3ex plus .2ex}{\large\bf}}

\def\eqnarray{\let\@currentlabel=\theequation\refstepcounter{equation}
    \global\@eqnswtrue
    \global\@eqcnt\z@\tabskip\@centering\let\\=\@eqncr
    $$\halign to \displaywidth\bgroup\@eqnsel\hskip\@centering
      $\displaystyle\tabskip\z@{##}$&\global\@eqcnt\@ne
       \hfil${{}##{}}$\hfil
      &\global\@eqcnt\tw@ $\displaystyle\tabskip\z@{##}$\hfil
       \tabskip\@centering&\llap{##}\tabskip\z@\cr}
\def\lefteqn#1{\hbox to 4\arraycolsep{$\displaystyle #1$\hss}}
\def\thesection{\arabic{section}}

\def\appendix{\setcounter{section}{0}
        \def\thesection{Appendix.}
        \def\theequation{\Alph{section}.\arabic{equation}}}

\long\def\@makefntext#1{\parindent 0cm\noindent
\hbox to 1em{\hss$^{\@thefnmark}$}#1}
\def\IR{{\hbox{{\rm I}\kern-.2em\hbox{\rm R}}}}
\def\IH{{\hbox{{\rm I}\kern-.2em\hbox{\rm H}}}}
\def\IC{{\ \hbox{{\rm I}\kern-.6em\hbox{\bf C}}}}
\def\IZ{{\hbox{{\rm Z}\kern-.4em\hbox{\rm Z}}}}
\def\rref#1{(\ref{#1})}
\newcommand{\beq}{\begin{equation}}
\newcommand{\eeq}{\end{equation}}
\newcommand{\ads}{$\widetilde{\hbox{adS}}$}
\newcommand{\NPB}[1]{{\sl Nucl.~Phys.}~{\bf B#1}}
\newcommand{\Ann}[1]{{\sl Ann.~Phys.}~{\bf #1}}
\newcommand{\CMP}[1]{{\sl Commun.~Math.~Phys.}~{\bf #1}}
\newcommand{\PLB}[1]{{\sl Phys.~Lett.}~{\bf B#1}}
\newcommand{\PLA}[1]{{\sl Phys.~Lett.}~{\bf A#1}}

\newcommand{\PRL}[1]{{\sl Phys.~Rev.\ Lett.}~{\bf #1}}

\newcommand{\MPLA}[1]{{\sl Mod.~Phys.\ Lett.}~{\bf A#1}}

\newcommand{\CQG}[1]{{\sl Class.~Quant.\ Grav.}~{\bf #1}}
\newcommand{\PRD}[1]{{\sl Phys.~Rev.}~{\bf D#1}}

\newcommand{\JMP}[1]{{\sl J.~Math.~Phys.}~{\bf #1}}
\newcommand{\GRG}[1]{{\sl Gen.~Rel.~Grav.}~{\bf #1}}
\begin{document}
%
%
%
%
\def\citen#1{%
\edef\@tempa{\@ignspaftercomma,#1, \@end, }
\edef\@tempa{\expandafter\@ignendcommas\@tempa\@end}%
\if@filesw \immediate \write \@auxout {\string \citation {\@tempa}}\fi
\@tempcntb\m@ne \let\@h@ld\relax \let\@citea\@empty
\@for \@citeb:=\@tempa\do {\@cmpresscites}%
\@h@ld}
%
\def\@ignspaftercomma#1, {\ifx\@end#1\@empty\else
   #1,\expandafter\@ignspaftercomma\fi}
\def\@ignendcommas,#1,\@end{#1}
%
%
\def\@cmpresscites{%
 \expandafter\let \expandafter\@B@citeB \csname b@\@citeb \endcsname
 \ifx\@B@citeB\relax 
    \@h@ld\@citea\@tempcntb\m@ne{\bf ?}%
    \@warning {Citation `\@citeb ' on page \thepage \space undefined}%
 \else
    \@tempcnta\@tempcntb \advance\@tempcnta\@ne
    \setbox\z@\hbox\bgroup 
    \ifnum\z@<0\@B@citeB \relax
       \egroup \@tempcntb\@B@citeB \relax
       \else \egroup \@tempcntb\m@ne \fi
    \ifnum\@tempcnta=\@tempcntb 
       \ifx\@h@ld\relax 
          \edef \@h@ld{\@citea\@B@citeB}%
       \else 
          \edef\@h@ld{\hbox{--}\penalty\@highpenalty \@B@citeB}%
       \fi
    \else   
       \@h@ld \@citea \@B@citeB \let\@h@ld\relax
 \fi\fi%
 \let\@citea\@citepunct
}
%
\def\@citepunct{,\penalty\@highpenalty\hskip.13em plus.1em minus.1em}%
%
%
\def\@citex[#1]#2{\@cite{\citen{#2}}{#1}}%
%
%
\def\@cite#1#2{\leavevmode\unskip
  \ifnum\lastpenalty=\z@ \penalty\@highpenalty \fi 
  \ [{\multiply\@highpenalty 3 #1
      \if@tempswa,\penalty\@highpenalty\ #2\fi 
    }]\spacefactor\@m}
\let\nocitecount\relax  
%
\begin{titlepage}
\vspace{.5in}
\begin{flushright}
UCD-95-15\\
gr-qc/9506079\\
May 1995\\
\end{flushright}
\vspace{.5in}
\begin{center}
{\Large\bf
 The (2+1)-Dimensional Black Hole}\\
\vspace{.4in}
{S.~C{\sc arlip}\footnote{\it email: carlip@dirac.ucdavis.edu}\\
       {\small\it Department of Physics}\\
       {\small\it University of California}\\
       {\small\it Davis, CA 95616}\\{\small\it USA}}
\end{center}

\vspace{.5in}
\begin{center}
\begin{minipage}{4.1in}
\begin{center}
{\large\bf Abstract}
\end{center}
{\small I review the classical and quantum properties of the
(2+1)-dimensional black hole of Ba{\~n}ados, Teitelboim, and
Zanelli.  This solution of the Einstein field equations in three
spacetime dimensions shares many of the characteristics of the
Kerr black hole: it has an event horizon, an inner horizon,
and an ergosphere; it occurs as an endpoint of gravitational
collapse; it exhibits mass inflation; and it has a nonvanishing
Hawking temperature and interesting thermodynamic properties.
At the same time, its structure is simple enough to allow a
number of exact computations, particularly in the quantum realm,
that are impractical in 3+1 dimensions.
}
\end{minipage}
\end{center}
\end{titlepage}
\addtocounter{footnote}{-1}

Since the seminal work of Deser, Jackiw, and 't~Hooft \cite{DJtH,
DJ,tH} and Witten \cite{Wita,Witb}, general relativity in three
spacetime dimensions has become an increasingly popular model in
which to explore the foundations of classical and quantum gravity
\cite{Korea}.  But although (2+1)-dimensional gravity has been
widely recognized as a useful laboratory for studying conceptual
issues---the nature of observables, for example, and the ``problem of
time''---it has been widely believed that the model is too physically
unrealistic to give much insight into real gravitating systems in
3+1 dimensions.  In particular, general relativity in 2+1 dimensions
has no Newtonian limit \cite{Barrow} and no propagating degrees of
freedom.

It therefore came as a considerable surprise when Ba{\~n}ados,
Teitelboim, and Zanelli (BTZ) showed in 1992 that (2+1)-dimensional
gravity has a black hole solution \cite{BTZ}.  The BTZ black hole
differs from the Schwarzschild and Kerr solutions in some important
respects: it is asymptotically anti-de Sitter rather than asymptotically
flat, and has no curvature singularity at the origin.  Nonetheless,
it is clearly a black hole: it has an event horizon and (in the
rotating case) an inner horizon, it appears as the final state of
collapsing matter, and it has thermodynamic properties much like
those of a (3+1)-dimensional black hole.

The purpose of this article is to briefly review the past three years'
work on the BTZ black hole.  The first four sections deal with classical
properties, while the last four discuss quantum mechanics, thermodynamics,
and possible generalizations.  For the most part, I will skip complicated
derivations, referring the reader instead to the literature.  For a
recent review with a somewhat complementary choice of topics, see
Ref.~\citen{Mannrev}.

The structure of the paper is as follows.  In section \ref{BH}, I
introduce the BTZ solution in standard Schwarzschild-like coordinates
and in Eddington-Finkelstein and Kruskal coordinates, and summarize
its basic physical characteristics.  Section \ref{GG} deals with the
global geometry of the (2+1)-dimensional black hole, and outlines
its description in the Chern-Simons formulation of (2+1)-dimensional
general relativity.  Section \ref{Col} describes the formation of BTZ
black holes from collapsing matter, and reports on some recent work on
critical phenomena, while section \ref{Int} summarizes the physics of
black hole interiors, focusing on the phenomenon of mass inflation.

I next turn to the quantum mechanical properties of the BTZ solution.
Section \ref{QFT} addresses the problem of quantum field theory in
a (classical) black hole background.  Sections \ref{Therm} and \ref{SM}
discuss the thermodynamic and statistical mechanical properties of the
quantized black hole, including attempts to explain black hole entropy
in terms of the ``microscopic'' physics of quantum gravitational states.
Finally, section \ref{Ass} briefly reviews a number of generalizations
of the BTZ solution, including electrically charged black holes,
dilatonic black holes, black holes in string theory, black holes in
topologically massive gravity, and black holes formed from ``topological''
matter.

\section{The BTZ Black Hole \label{BH}}
\setcounter{footnote}{0}

The BTZ black hole in ``Schwarzschild'' coordinates is described by
the metric
\beq
ds^2 = -( N^\perp)^2dt^2 + f^{-2}dr^2
  + r^2\left( d\phi + N^\phi dt\right)^2
\label{a1}
\eeq
with lapse and shift functions and radial metric
\beq
N^\perp = f
  = \left( -M + {r^2\over\ell^2} + {J^2\over4r^2} \right)^{1/2} ,
  \quad N^\phi = - {J\over2r^2} \qquad  (|J|\le M\ell) .
\label{a2}
\eeq
It is straightforward to check that this metric satisfies the
ordinary vacuum field equations of (2+1)-dimensional general
relativity,
\beq
R_{\mu\nu} - {1\over2} g_{\mu\nu}R = {1\over\ell^2}g_{\mu\nu}
\label{a3}
\eeq
with a cosmological constant $\Lambda=-1/\ell^2$.  The metric \rref{a1}
is stationary and axially symmetric, with Killing vectors $\partial_t$
and $\partial_\phi$, and generically has no other symmetries.

As the notation suggests, the parameters $M$ and $J$, which determine
the asymptotic behavior of the solution, are the standard ADM mass
and angular momentum.  To see this, one can write the Einstein action
in ADM form,\footnote{Unless otherwise stated, I use the units of
Ref.~\citen{BTZ}, in which $8G=1$.}
\beq
I = {1\over2\pi}\int_0^T \!dt\int\!d^2x\,\left[\pi^{ij}\dot g_{ij}
  - N^\perp{\cal H} - N^i{\cal H}_i\right] + B ,
\label{a4}
\eeq
where the boundary term $B$ is required to cancel surface integrals
in the variation of $I$ and ensure that the action has genuine extrema
\cite{RegTeit}.  If one now considers variations of the spatial metric
that preserve the asymptotic form of the BTZ solution, one finds that
\cite{BHTZ}
\beq
\delta B = T\left[ -\delta M + N^\phi\delta J \right].
\label{a5}
\eeq
As in 3+1 dimensions, the conserved charges can be read off from
$\delta B$: $J$ is the angular momentum as measured at infinity, while
$M$ is the mass associated with asymptotic translations in the ``Killing
time'' $t$.

This analysis can be formalized by noting that $M$ and $J$
are the Noether charges, as defined by Lee and Wald \cite{Wald},
associated with asymptotic time translations and rotations (see
\cite{CGeg} for a computation in the first-order formalism).
It may also be checked that $M$ and $J$ are the conserved charges
associated with the asymptotic Killing vectors $\partial_t$ and
$\partial_\phi$ as defined by Abbott and Deser \cite{Abbott} for
asymptotically anti-de Sitter spacetimes, and that they can be obtained
from the anti-de Sitter version of the stress-energy pseudotensor
considered by Bak et al.\ \cite{Bak}.  Alternatively, $M$ and $J$ may
be expressed in terms of the quasilocal energy and angular momentum of
Brown and York \cite{BYork}: if one places the BTZ black hole in a
circular box of radius $R$ and treats the boundary terms at $R$
carefully, one obtains a mass and angular momentum measured at the
boundary that approach $M$ and $J$ as $R\rightarrow\infty$ \cite{BMann,
Mannrev}.

The metric \rref{a1} is singular when $r\!=\!r_\pm$, where
\beq
r_\pm^2={M\ell^2\over 2}\left \{ 1 \pm
\left [ 1 - \left({J\over M\ell}\right )^2\right ]^{1/2}\right \} ,
\label{a6}
\eeq
i.e.,
\beq
M={r_+^2+r_-^2\over\ell^2}, \quad J={2r_+ r_-\over\ell} \ .
\label{a6a}
\eeq
As we shall see below, these are merely coordinate singularities, closely
analogous to the singularity at $r\!=\!2m$ of the ordinary Schwarzschild
metric.  The time-time component $g_{00}$ of the metric vanishes at
$r\!=\!r_{\hbox{\scriptsize\em erg}}$, where
\beq
r_{\hbox{\scriptsize\em erg}} = M^{1/2}\ell
  = \left(r_+^2+r_-^2\right)^{1/2} .
\label{a7}
\eeq
As in the Kerr solution in 3+1 dimensions, $r\!<\!r_{\hbox{\scriptsize
\em erg}}$ determines an ergosphere:  timelike curves in this region
necessarily have $d\phi/d\tau\!>\!0$ (when $J\!>\!0$), so all observers
are dragged along by the rotation of the black hole.  Note that the
$r_\pm$ become complex if $|J|\!>\!M\ell$, and the horizons disappear,
leaving a metric that has a naked conical singularity at $r\!=\!0$.  The
$M\!=\!-1$, $J\!=\!0$ metric may be recognized as that of ordinary
anti-de Sitter space; it is separated by a mass gap from the $M\!=\!0$,
$J\!=\!0$ ``massless black hole,'' whose geometry is discussed in
Refs.~\citen{BHTZ} and \citen{Steif2}.

That the BTZ metric is a genuine black hole can be seen most easily by
transforming to Eddington-Finkelstein-like coordinates
\cite{ChanMann},
\beq
dv = dt + {dr\over (N^\perp)^2} , \quad
d\tilde\phi = d\phi - {N^\phi\over (N^\perp)^2} dr ,
\label{a8}
\eeq
in which the metric becomes
\beq
ds^2 = - (N^\perp)^2 dv^2 + 2dvdr
  + r^2\left( d\tilde\phi + N^\phi dv\right)^2 .
\label{a9}
\eeq
It is now easy to see that the horizon $r=r_+$, where $N^\perp$ vanishes,
is a null surface, generated by geodesics
\beq
r(\lambda)=r_+ , \quad
{d\tilde\phi\over d\lambda} + N^\phi(r_+){dv\over d\lambda} = 0 .
\label{a10}
\eeq
Moreover, this surface is evidently a marginally trapped surface: at
$r\!=\!r_+$, any null geodesic satisfies
\beq
{dv\over d\lambda}{dr\over d\lambda} = -{r_+^2\over2}
\left({d\tilde\phi\over d\lambda} + N^\phi(r_+){dv\over d\lambda}\right)^2
\le 0 ,
\label{a10a}
\eeq
so $r$ decreases or (for the geodesics \rref{a10}) remains constant as
$v$ increases.

Like the outer horizon of the Kerr metric, the surface $r\!=\!r_+$ is
also a Killing horizon.  The Killing vector normal to this surface is
\beq
\chi = \partial_v - N^\phi(r_+)\partial_{\tilde\phi} ,
\label{a11}
\eeq
from which the surface gravity $\kappa$, defined by \cite{Wald2}
\beq
\kappa^2 = -{1\over2}\nabla^a\chi^b\nabla_a\chi_b ,
\eeq
may be computed to be
\beq
\kappa = {r_+^2-r_-^2\over\ell^2r_+} .
\label{a12}
\eeq

For a more complete description of the BTZ solution, we can transform
instead to Kruskal-like coordinates \cite{BHTZ}.  To do so, let us
define new null coordinates
\beq
u = \rho(r)e^{-at} ,\quad \quad v = \rho(r)e^{at} ,
\qquad\hbox{with}\quad {d\rho\over dr} = {a\rho\over (N^\perp)^2} .
\label{a13}
\eeq
As in the case of the Kerr metric, we need two patches, $r_-<r<\infty$
and $0<r<r_+$, to cover the BTZ spacetime.  In each patch, the metric
\rref{a1} takes the form
\beq
ds^2 = \Omega^2 dudv + r^2 (d\tilde\phi + N^\phi dt)^2
\label{a14}
\eeq
where
\begin{eqnarray}
\Omega_+^2 &=& {(r^2-r_-^2)(r+r_+)^2\over a_+^2r^2\ell^2}
  \left({r-r_-\over r+r_-}\right)^{r_-/r_+} , \nonumber\\
&&\tilde\phi_+ = \phi + N^\phi(r_+)t\, , \qquad
a_+ = {r_+^2-r_-^2\over\ell^2r_+} \qquad\qquad   (r_-<r<\infty) \\
\Omega_-^2 &=& {(r_+^2-r^2)(r+r_-)^2\over a_-^2r^2\ell^2}
  \left({r_+-r\over r_++r}\right)^{r_+/r_-} , \nonumber\\
&&\tilde\phi_- = \phi + N^\phi(r_-)t\, , \qquad
a_- = {r_-^2-r_+^2\over\ell^2r_-} \qquad\qquad   (0<r<r_+)
\label{a15}
\end{eqnarray}
with $r$ and $t$ viewed as implicit functions of $u$ and $v$.  (For
explicit coordinate transformations, see Ref.~\citen{BHTZ}.\footnote{Note
that the coordinates $U$ and $V$ in \cite{BHTZ} are, unconventionally,
not null; my $u$ and $v$ are the sum and difference of the coordinates
of this reference.})

As in the case of the Kerr black hole, an infinite number of such Kruskal
patches may be joined together to form a maximal solution, whose Penrose
diagram is shown in figure 1a.  This diagram differs from that of the
Kerr metric at $r\!=\!\infty$, reflecting the fact that the BTZ black hole
is asymptotically anti-de Sitter rather than asymptotically flat, but
the overall structure is similar.  In particular, it is evident that
$r\!=\!r_+$ is an event horizon, while the inner horizon $r\!=\!r_-$ is
a Cauchy horizon for region I.  When $J\!=\!0$, the Penrose diagram
collapses to that of figure 1b, which is similar in structure---except
for its asymptotic behavior---to the diagram for the ordinary Schwarzschild
solution, while for the extreme case, $J\!=\!\pm M\ell$, the Penrose
diagram is that of figure 1c.

Although the classical behavior of the BTZ black hole has not been
investigated as thoroughly as, for example, the Schwarzschild solution,
a fair amount is known.  In particular, the behavior of geodesics has
been studied by Cruz et al.\ \cite{Cruz} and Farina et al.\ \cite{Farina},
and the propagation of strings in a BTZ background has been analyzed by
Larsen and Sanchez \cite{LarSan}.  The static BTZ black hole has also
been studied in the York time slicing, in which surfaces of constant
mean (extrinsic) curvature $\hbox{Tr}K=T$ are used as constant time
surfaces; the resulting metric is equivalent to \rref{a1}, via a
complicated coordinate transformation involving elliptic integrals
\cite{CarFuHos}.  There have also been some recent attempts to find
exact multi-black hole solutions.  Cl{\'e}ment has reported the existence
of solutions representing multiple freely falling black holes, but the
metrics also contain conical singularities \cite{Clement}.  Coussaert
and Henneaux claim that all static multi-black hole solutions contain
such singularities \cite{CousHenn}, but it is not known whether exact
nonstatic nonsingular solutions can be found.

\section{Global Geometry \label{GG}}
\setcounter{footnote}{0}

In the last section, I emphasized similarities between the BTZ solution
and ordinary (3+1)-dimensional black holes.  But there are important
differences as well, rooted in the simplicity of (2+1)-dimensional
gravity.  In three spacetime dimensions, the full curvature tensor is
completely determined by the Ricci tensor,
\beq
R_{\mu\nu\rho\sigma} = g_{\mu\rho}R_{\nu\sigma}
+ g_{\nu\sigma}R_{\mu\rho} - g_{\nu\rho}R_{\mu\sigma}
- g_{\mu\sigma}R_{\nu\rho} - {1\over2}
(g_{\mu\rho}g_{\nu\sigma} - g_{\mu\sigma}g_{\nu\rho})R .
\label{b1}
\eeq
Hence any solution of the vacuum Einstein field equations with a
cosmological constant $\Lambda$,
\beq
R_{\mu\nu} = 2\Lambda g_{\mu\nu} ,
\label{b2}
\eeq
has constant curvature.  In particular, the BTZ metric \rref{a1} has
constant  negative curvature: any point in the black hole spacetime
has a neighborhood isometric to anti-de Sitter space, and the whole
spacetime is expressible as a collection of such neighborhoods
appropriately patched together.  Since the maximal simply connected
spacetime of constant negative curvature is the universal covering
space \ads\ of anti-de Sitter space, we might hope to represent the
BTZ black hole as a quotient space of \ads\  by some group of
isometries.\footnote{For certain spacetime topologies, the general
existence of such a quotient space construction has been proven
by Mess \cite{Mess}.}  Such a quotient construction provides a
powerful mathematical tool, permitting, for example, the exact
computation of Greens functions in a black hole background.

Geometrically, three-dimensional anti-de Sitter space (adS) may
be obtained from flat $\IR^{2,2}$, with coordinates $(X_1,X_2,T_1,T_2)$
and metric
\beq
dS^2 = dX_1^2 + dX_2^2 - dT_1^2 - dT_2^2 ,
\label{b2a}
\eeq
by restricting to the submanifold
\beq
X_1^2-T_1^2+X_2^2-T_2^2 = -\ell^2
\label{b3}
\eeq
with the induced metric.  In this formulation, the isometry group
is evidently $\hbox{SO}(2,2)$.  Equivalently, we can combine
$(X_1,X_2,T_1,T_2)$ into a $2\times2$ matrix,
\beq
{\bf X} = {1\over\ell}\left( \begin{array}{cc}
  T_1+X_1 & T_2+X_2\\ -T_2+X_2 & T_1-X_1 \end{array} \right) ,
\qquad \hbox{det}|{\bf X}| = 1 ,
\label{b4}
\eeq
i.e., ${\bf X}\!\in\!\hbox{SL}(2,\IR)$.  Isometries may now be represented
as elements of the group $\hbox{SL}(2,\IR)\times\hbox{SL}(2,\IR)/\IZ_2\!
\approx\!\hbox{SO}(2,2)$: the two copies of $\hbox{SL}(2,\IR)$ act by left
and right multiplication, ${\bf X}\!\rightarrow\!\rho_L{\bf X}\rho_R$, with
$(\rho_L,\rho_R)\!\sim\!(-\rho_L,-\rho_R)$.

The relevant region of the universal covering space of anti-de Sitter
space (see Ref.~\citen{BHTZ}) may be covered by an infinite set of
coordinate patches of three types, corresponding to the regions of the
Penrose diagram of figure 1a:
\begin{description}
\item[I.\ ($r\ge r_+$)\\]
\begin{eqnarray}
X_1=\ell\sqrt{\alpha}
  \sinh\left({r_+\over\ell}\phi-{r_-\over\ell^2}t\right) &,&\quad\!\!
X_2=\ell\sqrt{\alpha-1}
  \cosh\left({r_+\over\ell^2}t-{r_-\over\ell}\phi\right) \nonumber\\
T_1=\ell\sqrt{\alpha}
   \cosh\left({r_+\over\ell}\phi-{r_-\over\ell^2}t\right) &,&\quad\!
T_2=\ell\sqrt{\alpha-1}
  \sinh\left({r_+\over\ell^2}t-{r_-\over\ell}\phi\right)  \nonumber
\label{b5}
\end{eqnarray}
\item[II.\ ($r_-\le r\le r_+$)\\]
\begin{eqnarray}
X_1=\ell\sqrt{\alpha}
  \sinh\left({r_+\over\ell}\phi-{r_-\over\ell^2}t\right) &,&\quad\!\!
X_2=-\ell\sqrt{1-\alpha}
  \sinh\left({r_+\over\ell^2}t-{r_-\over\ell}\phi\right) \nonumber\\
T_1=\ell\sqrt{\alpha}
   \cosh\left({r_+\over\ell}\phi-{r_-\over\ell^2}t\right) &,&\quad\!
T_2=-\ell\sqrt{1-\alpha}
  \cosh\left({r_+\over\ell^2}t-{r_-\over\ell}\phi\right)  \nonumber
\end{eqnarray}
\item[III.\ ($0\le r\le r_-$)\\]
\begin{eqnarray}
X_1=\ell\sqrt{-\alpha}
  \cosh\left({r_+\over\ell}\phi-{r_-\over\ell^2}t\right) &,&\quad\!\!
X_2=-\ell\sqrt{1-\alpha}
  \sinh\left({r_+\over\ell^2}t-{r_-\over\ell}\phi\right) \nonumber\\
T_1=\ell\sqrt{-\alpha}
   \sinh\left({r_+\over\ell}\phi-{r_-\over\ell^2}t\right) &,&\quad\!
T_2=-\ell\sqrt{1-\alpha}
  \cosh\left({r_+\over\ell^2}t-{r_-\over\ell}\phi\right)  \nonumber\\
\end{eqnarray}
\end{description}
where
\beq
\alpha(r)=\left({r^2-r_-^2\over r_+^2-r_-^2}\right),
  \quad \phi\in(-\infty,\infty),\ t\in(-\infty,\infty) .
\label{b8}
\eeq
It is straightforward to show that the standard adS metric $dS^2$
then transforms to the BTZ metric \rref{a1} in each patch.  The ``angle''
$\phi$ in equation \rref{b5} has infinite range, however; to make it
into a true angular variable, we must identify $\phi$ with $\phi+2\pi$.
This identification is an isometry of anti-de Sitter space---it is
a boost in the $X_1$-$T_1$ and the $X_2$-$T_2$ planes---and
corresponds to an element $(\rho_L,\rho_R)$ of $\hbox{SL}(2,\IR)
\times\hbox{SL}(2,\IR)/\IZ_2$ with
\beq
\rho_L=\left( \begin{array}{cc}
 e^{{\pi (r_+-r_-)/\ell}} & 0 \\ 0 & e^{-{\pi (r_+-r_-)/\ell}}
 \end{array}\right) ,\qquad
\rho_R=\left( \begin{array}{cc}
 e^{{\pi (r_++r_-)/\ell}} & 0 \\ 0 & e^{-{\pi (r_++r_-)/\ell}}
 \end{array}\right) .
\label{b9}
\eeq
The BTZ black hole may thus be viewed as a quotient space
$\widetilde{\hbox{adS}}/\langle(\rho_L,\rho_R)\rangle$, where
$\langle(\rho_L,\rho_R)\rangle$ denotes the group generated by
$(\rho_L,\rho_R)$.  This is an extraordinary result: anti-de Sitter
space is an extremely simple, virtually structureless manifold,
but appropriate identifications nevertheless convert it into a
spacetime very much like the (3+1)-dimensional Kerr black hole.

A slightly different representation for the region $r\ge r_+$ will be
useful for investigating the first-order formulation of the BTZ black
hole.  This ``upper half-space'' metric may be obtained from \rref{a1}
by the coordinate transformation
\begin{eqnarray}
x &=& \left({r^2-r_+^2\over r^2-r_-^2}\right)^{1/2}
      \cosh\left( {r_+\over\ell^2}t - {r_-\over\ell}\phi \right)
      \exp\left\{ {r_+\over\ell}\phi - {r_-\over\ell^2}t \right\}
      \nonumber \\
y &=& \left({r^2-r_+^2\over r^2-r_-^2}\right)^{1/2}
      \sinh\left( {r_+\over\ell^2}t - {r_-\over\ell}\phi \right)
      \exp\left\{ {r_+\over\ell}\phi - {r_-\over\ell^2}t \right\} \\
z &=& \left({r_+^2-r_-^2\over r^2-r_-^2}\right)^{1/2}
      \exp\left\{ {r_+\over\ell}\phi - {r_-\over\ell^2}t \right\} ,
      \nonumber
\label{b10}
\end{eqnarray}
for which the BTZ metric becomes
\beq
ds^2 = {\ell^2\over z^2} (dx^2 - dy^2 + dz^2) \qquad (z>0) .
\label{b11}
\eeq
Again, periodicity in the Schwarzschild angular coordinate $\phi$
requires that we identify points under the action $\phi\rightarrow
\phi+2\pi$, that is,
\begin{eqnarray}
&(&x,y,z)\sim \\
&\phantom{.}&\left(
e^{2\pi r_+/\ell}(x\cosh{2\pi r_-\over\ell} - y\sinh{2\pi r_-\over\ell}),\,
e^{2\pi r_+/\ell}(y\cosh{2\pi r_-\over\ell} - x\sinh{2\pi r_-\over\ell}),\,
e^{2\pi r_+/\ell}z\right) \nonumber .
\label{b12}
\end{eqnarray}
These identifications correspond once more to the isometry \rref{b9}
of \ads.

The physical significance of the quotient space representation of
the BTZ black hole may be clarified by turning to the first-order
formulation of general relativity.  As discovered by Ach\'ucarro
and Townsend \cite{Achu} and developed by Witten \cite{Wita,Witb},
(2+1)-dimensional gravity can be rewritten as a Chern-Simons gauge theory.
The fundamental variables in the first-order formalism are a triad
$e^a\!=\!e_\mu{}^adx^\mu$ (where $g_{\mu\nu}\!=\!\eta_{ab}e_\mu{}^a
e_\nu{}^b$) and a spin connection $\omega^a\!=\!{1\over2}\epsilon^{abc}
\omega_{\mu bc}dx^\mu$, which can be combined to give a pair of
$\hbox{SO}(2,1)$ or $\hbox{SL}(2,\IR)$ connection one-forms
\beq
A^{(\pm)a} = \omega^a \pm {1\over\ell} e^a .
\label{b13}
\eeq
It is not hard to show that the Einstein-Hilbert action written in terms
of these ``gauge fields'' becomes
\beq
I_{\hbox{\scriptsize grav}} =
  I_{\hbox{\scriptsize CS}}[A^{(+)}] -I_{\hbox{\scriptsize CS}}[A^{(-)}] ,
\label{b14}
\eeq
where
\beq
I_{\hbox{\scriptsize CS}} = {k\over4\pi}\int_M\,\hbox{Tr}
  \left\{ A\wedge dA + {2\over3}A\wedge A\wedge A\right\} ,
\label{b15}
\eeq
is the Chern-Simons action.  (Here $k$ is a constant proportional to
$\ell/G$; its exact value depends on normalizations, and will not be
important until section \ref{SM}.)  The Chern-Simons field equations
require that the connections $A^{(\pm)}$ be flat, and it may be checked
that this is equivalent to the constant curvature condition \rref{b1}.

Now, any connection is completely determined by its holonomies,
that is, by the Wilson loops
\beq
H[\gamma] = P\exp\left\{\int_\gamma A\right\}
\label{b16}
\eeq
around closed curves $\gamma$, where $P$ denotes path ordering.  In
3+1 dimensions, such holonomies are the fundamental variables in
the ``loop representation'' \cite{RovSmo}, and in 2+1 dimensions they
play an equally important role.\footnote{For a general discussion of
the geometrical significance of these holonomies in 2+1 dimensions, see
Ref.~\citen{Carhol}.}  For a flat connection, the holonomies depend only
on the homotopy class of $\gamma$, and may be thought of as non-Abelian
Aharonov-Bohm phases.  The $H[\gamma]$ are not quite gauge invariant,
but transform by overall conjugation, $H[\gamma]\!\rightarrow\!g
\cdot H[\gamma]\cdot g^{-1}$; their traces give a gauge-invariant
and diffeomorphism-invariant characterization of the geometry.

For the metric \rref{b10}--\rref{b11} with the identifications
\rref{b12}, the closed curve
\beq
\gamma: \phi(s) = 2\pi s , \quad s\in[0,1]
\label{b16a}
\eeq
is homotopically nontrivial, and it may be shown that up to overall
conjugation, the holonomies of the connections $A^{(\pm)}$ are precisely
the elements $\rho_L$ and $\rho_R$ given by equation \rref{b9}.  The
holonomies thus unite the gauge theoretic and the geometric descriptions
of the black hole---the same group elements that determine the Chern-Simons
connection also give the set of identifications that fix the geometry.
(See also Vaz and Witten \cite{Vaz} for a discussion of the geometric
interpretation of the $H[\gamma]$.)  An equivalent expression for the
holonomies in the ``Schwarzschild'' coordinates \rref{a1} was first
discovered by Cangemi et al. \cite{Cangemi}, who pointed out that the
mass and angular momentum \rref{a6a} have a natural interpretation in
terms of the two quadratic Casimir operators of $\hbox{SL}(2,\IR)\!
\times\!\hbox{SL}(2,\IR)$.  This connection between mass, angular
momentum, and holonomies has also been investigated by Izquierdo and
Townsend \cite{IzTown}, who discuss the possibility that bounds on
the holonomies might give a version of the positive mass theorem for
nonsingular (2+1)-dimensional asymptotically anti-de Sitter spacetimes.

The quotient space construction of the BTZ black hole has a number
of uses.  For example, we can now investigate the nature of the
singularity at $r\!=\!0$ more easily.  Note first that there is no
curvature singularity: from \rref{b1}, the curvature is constant
everywhere.  Moreover, it would seem from \rref{b5} that the metric
could be extended in region III past $r\!=\!0$ to negative values of
$r^2$.  This is indeed possible, but it is shown in Ref.~\citen{BHTZ}
that the resulting manifold contains closed timelike curves: the Killing
vector describing the identifications $\langle(\rho_L,\rho_R)\rangle$
of \ads\  becomes timelike in these regions, so curves such as \rref{b16a}
become timelike.  The surface $r\!=\!0$ is thus a singularity in the
causal structure.\footnote{For $J\!=\!0$, the spacetime fails to be
Hausdorff at $r=0$, and the singularity resembles that of Taub-NUT
space, but this is not the case if $J\!\ne\!0$; see appendix B of
Ref.~\citen{BHTZ}.}

The quotient space construction is also useful for identifying Killing
spinors.  Such spinors are solutions of the equation
\beq
\nabla_\lambda \psi = {\epsilon\over2\ell}\gamma_\lambda\psi ,
\label{b17}
\eeq
where $\epsilon\!=\!\pm 1$ \cite{CousHenn}; the choice of sign of
$\epsilon$ may be viewed as a choice of which of the two factors
of $\hbox{SL}(2,\IR)$ in the isometry group to gauge.  The existence
of Killing spinors is an indication of supersymmetry: if one views the
BTZ black hole as a solution of (1,1)-adS supergravity with vanishing
gravitino fields, a Killing spinor represents a remaining supersymmetry
transformation that leaves this metric and gravitino configuration
invariant.

The Killing spinors of anti-de Sitter space are easy to find---there
are four, two for each sign of $\epsilon$---and their existence in
the BTZ spacetime is simply a question of whether the adS solutions
are preserved by the identifications $\langle(\rho_L,\rho_R)\rangle$.
In general, they are not: as Coussaert and Henneaux have shown, Killing
spinors exist only for extreme black holes, those for which $J\!=\!\pm
M\ell$ \cite{CousHenn}.  When $M\!\ne\!0$, an extreme black hole has
a single supersymmetry; the $M\!=\!0$ black hole has two, each periodic
in the angular coordinate $\phi$, and may be viewed as the ground state
of the Ramond sector of (1,1)-adS supergravity.  Steif has recently
shown that these Killing spinors have an even more direct geometric
interpretation \cite{Steif2}: anti-de Sitter space may be embedded in
the group manifold of the supergroup $\hbox{OSp}(1|2;\IR)$, whose
isometry group is $\hbox{OSp}(1|2;\IR)\!\times\!\hbox{OSp}(1|2;\IR)/\IZ_2$,
and the Killing spinors are simply the generators of odd elements of
this group that commute with the even elements $(\rho_L,\rho_R)$.
Supersymmetric solutions of (2,0)-adS supergravity have also been
investigated \cite{IzTown}; a number of supersymmetric configurations
exist, but they typically involve naked conical singularities.

\section{Black Holes and Gravitational Collapse \label{Col}}
\setcounter{footnote}{0}

The physical importance of the (3+1)-dimensional black hole comes
from its role as the final state of gravitational collapse.  Since
(2+1)-dimensional gravity has no Newtonian limit, one might fear that
no such interpretation exists for the BTZ black hole. In fact, however,
it was shown shortly after the discovery of the BTZ solution that this
black hole arises naturally from collapsing matter \cite{MannRoss}.

Consider a (2+1)-dimensional spacetime containing a spherical cloud of
dust surrounded by empty space.  For the exterior, we take the metric
to be of the BTZ form \rref{a1}--\rref{a2}, with $J=0$ for simplicity;
for the interior, we may choose comoving coordinates, in which the
geometry is given by a Robertson-Walker metric,
\beq
ds^2 = -dt^2 + a^2(t)\left( {d\tilde r^2\over 1-k\tilde r^2}
  + \tilde r^2 d\phi^2 \right).
\label{c1}
\eeq
The stress-energy tensor for pressureless dust is
\beq
T_{\mu\nu} = \rho(t) u_\mu u_\nu ,
\label{c2}
\eeq
with $u_\mu=(1,0,0)$ in comoving coordinates; as usual, conservation
implies that $\rho a^2=\rho_0a_0{}^2$.  It is then easy to show that the
field equations in the interior are solved by
\begin{eqnarray}
a(t)&=& a_0\cos{t\over\ell} + \ell{\dot a}_0\sin{t\over\ell} ,
\nonumber\\
{\dot a}_0{}^2&=& 8\pi G\rho_0a_0{}^2 - k -{a_0{}^2\over\ell^2} .
\label{c3}
\end{eqnarray}
Note that for arbitrary initial values, $a(t)$ always reaches zero in
a finite proper time.

We must now join the interior and exterior solutions, using the
standard matching conditions that the spatial metric $g_{ij}$ and the
extrinsic curvature $K_{ij}$ be continuous at the boundary.  As Mann
and Ross show \cite{MannRoss}, this requires that
\beq
M = 8\pi G\rho_0a_0{}^2r_0{}^2 - 1 ,
\label{c4}
\eeq
where $\tilde r\!=\!r_0$ is the position of the surface of the collapsing
dust in the interior (comoving) coordinate, equivalent to the exterior
radial coordinate $r=r_0a(t)$.  The collapse closely resembles the
Oppenheimer-Snyder solution in 3+1 dimensions.  In particular, the
surface of the collapsing dust crosses the horizon in a finite amount
of comoving time, but light emitted from the surface is infinitely
red-shifted at the horizon, and the collapse appears to take infinitely
long to a static exterior observer.

The mass $M$ of the final black hole depends on three parameters,
$\rho_0$, $r_0$, and $a_0$, or equivalently $\rho_0$, $r_0$, and
$v_0$, where $v_0\!=\!{\dot r}|_{t=0}\!=\!r_0{\dot a}_0$ is the
initial velocity.  If these parameters are such that $M\!<\!0$ in
eqn.~\rref{c4}, the final state is not a black hole, but rather a
naked conical singularity in an asymptotically anti-de Sitter spacetime.
We have thus found a sort of ``phase transition'' in the space of
initial values, quite similar to the transition discovered numerically
by Choptuik in 3+1 dimensions \cite{Choptuik}.  This transition has
been investigated in detail by Peleg and Steif for the related case
of a collapsing thin shell of dust \cite{Peleg}.  In that case, there
are four possible final states: open conical adS space, the BTZ black
hole interior, the BTZ black hole exterior, and closed conical adS
space.  A phase diagram is shown in figure 2.  Peleg and Steif show
that the transitions between black hole configurations and naked
singularities are critical phenomena characterized by an order
parameter with critical exponent $1/2$, again mimicking the behavior
of more complicated (3+1)-dimensional models (although with a different
critical exponent).

While pressureless dust in 2+1 dimensions necessarily collapses
(provided that $\Lambda<0$), a (2+1)-dimensional ball of fluid can
be stabilized by internal pressure.  The general properties of such
(2+1)-dimensional ``stars'' have recently been investigated by
Cruz and Zanelli \cite{CruzZan}.  They find that for a wide range
of equations of state, a static interior solution can be joined to
a BTZ exterior solution.  They also find an upper limit for the
mass of a stable star of fixed radius, but the result depends on
an integration constant whose physical significance is not completely
clear.

A BTZ black hole can also form from a collapsing pulse of radiation
\cite{Husain,Virb,ChanMann}.  The most useful coordinates to describe
this process are the Eddington-Finkelstein coordinates of \rref{a9},
in which a suitable metric ansatz---analogous to the Vaidya metric in
3+1 dimensions---is
\beq
ds^2 = \left[{r^2\over\ell^2} + m(v)\right] dv^2
  + 2dvdr - j(v)dvd\tilde\phi + r^2 d\tilde\phi^2 .
\label{c5}
\eeq
For the stress-energy tensor, we take that of a rotating null fluid,
\beq
T_{vv} = {\rho(v)\over r} + {j(v)\,\omega(v)\over2r^3} , \quad
T_{v\tilde\phi} = -{\omega(v)\over r} ,
\label{c6}
\eeq
where $\rho(v)$ and $\omega(v)$ are arbitrary functions and the form
of the $r$ dependence follows from the conservation law for the
stress-energy tensor.  The Einstein field equations
then reduce to
\beq
{dm(v)\over dv} = 2\pi\rho(v) , \qquad
{d\,j(v)\over dv} = 2\pi\omega(v)
\label{c7}
\eeq
(in BTZ units, $8G\!=\!1$).  In particular, any distribution of radiation
for which $m(v)\!\sim\!M$ and $j(v)\!\sim\!J$ as $v$ goes to infinity
will approach a BTZ black hole.  The pulse of radiation considered by
Husain \cite{Husain}, for instance,
\beq
\rho(v) = A\,\hbox{sech}^2{v\over b} ,
\label{c8}
\eeq
leads asymptotically to a black hole with mass $2\pi A$.

\section{Interiors \label{Int}}
\setcounter{footnote}{0}

One of the important open questions in black hole physics is that of
the stability of the inner horizon.  It is apparent from the Penrose
diagram of figure 1 that in the case of a rotating black hole, an
infalling observer need not hit the singularity at $r\!=\!0$, but can
escape through the inner horizon $r\!=\!r_-$ to a new exterior region.
On the other hand, infalling radiation is infinitely blue-shifted at
$r\!=\!r_-$, suggesting that the inner horizon is not stable; and
indeed, simple models in 3+1 dimensions indicate that this horizon
may be destroyed by the back reaction of ingoing and back-scattered
outgoing radiation.  This phenomenon has been an important focus of
research in (3+1)-dimensional general relativity \cite{Poisson,Ori},
where it has been shown that the internal mass function of a
Reissner-Nordstrom black hole---or equivalently, the ``Coulomb''
component $|\Psi_2|$ of the Weyl tensor---diverges at the Cauchy
horizon (``mass inflation''), but that tidal forces may nevertheless
remain weak enough that physical objects can survive passage through
the horizon.  (For a nice review, see Ref.~\citen{Israel}.)

In 2+1 dimensions, an exact computation analogous to that of
Refs.~\citen{Poisson}--\citen{Ori} may be performed for the rotating
BTZ black hole \cite{ChanMann,Husain}.  Since the resulting phenomenon
of mass inflation is discussed in some detail in Ref.~\citen{Mannrev},
I will only briefly summarize the results.

The starting point is again the metric \rref{c5}.  More precisely,
let us model a thin shell of {\em outgoing} radiation by joining an
``exterior'' metric of the form \rref{c5}, with mass function $m_1(v_1)$,
to an ``interior'' metric, also of the form \rref{c5}, with a mass
function $m_2(v_2)$.  By choosing appropriate matching conditions for
the interior and exterior regions, we can model the interaction of
infalling radiation---$\rho(v)$ in eqn.~\rref{c6}---with this shell of
outgoing radiation.  A careful analysis of these matching conditions
then shows that $m_2$ necessarily diverges at the Cauchy horizon.
(For the spinning case, it is actually the quantity
\beq
E_2(v) = m_2(v) - j_2{}^2(v)/4r^2
\label{d1}
\eeq
that diverges.)  The (2+1)-dimensional black hole thus exhibits mass
inflation.  On the other hand, as in 3+1 dimensions, tidal forces lead
to only a finite distortion at the Cauchy horizon, so it is not clear
that passage through the horizon is forbidden.  Moreover, in contrast to
the (3+1)-dimensional case, the Kretschmann scalar $R_{\mu\nu\rho\sigma}
R^{\mu\nu\rho\sigma}$ remains finite.  Further investigation of more
complicated solutions---in particular, solutions with realistic outgoing
radiation---seems feasible, and could provide valuable information on the
question of stability.

\section{Quantum Field Theory in a Black Hole Background \label{QFT}}
\setcounter{footnote}{0}

While the BTZ black hole is useful as a comparatively simple model for
classical black hole physics, its real power appears when we turn to
quantum theory.  General relativity in 2+1 dimensions has proven to
be an instructive model for realistic (3+1)-dimensional quantum gravity
in a number of settings \cite{Korea}, and the black hole is no exception.

As in 3+1 dimensions, it is useful to warm up by considering quantum
field theory in a classical black hole background.  For a free field,
the starting point for such a theory is an appropriate two-point
function $G(x,x')\!=\!\langle0|\phi(x)\phi(x')|0\rangle$, from which
such quantities as the expectation values $\langle0|T_{\mu\nu}|0\rangle$
can be derived.  The key simplification in 2+1 dimensions comes from
the quotient space picture of the BTZ black hole described in section
\ref{GG}, that is, its representation as a region of the universal
covering space of anti-de Sitter space with appropriate identifications.
This construction allows us to write the two-point function for the
black hole in terms of the corresponding \ads\  two-point function by
means of the method of images.  Specifically, if $G_A(x,x')$ is a
two-point function in \ads, the corresponding function for the BTZ
black hole is
\beq
G_{BTZ}(x,x') = \sum_n e^{-i\delta n} G_A(x,\Lambda^nx') ,
\label{e1}
\eeq
where $\Lambda x'$ denotes the action of the group element \rref{b9}
on $x'$.  The phase $\delta$ is zero for ordinary (``untwisted'')
fields, but may in principle be arbitrary, corresponding to boundary
conditions $\phi(\Lambda x)\!=\!e^{-i\delta}\phi(x)$; the choice $\delta
\!=\!\pi$ leads to conventional ``twisted'' fields.  Our problem
has thus been effectively reduced to the comparatively simple problem
of understanding quantum field theory on \ads.

While quantum field theory on anti-de Sitter space is fairly simple,
it is by no means trivial.  The main difficulty comes from the fact
that neither anti-de Sitter space nor its universal covering space
are globally hyperbolic.  As is evident from the Penrose diagrams of
figure 1, spatial infinity is timelike, and information may enter or
exit from the ``boundary'' at infinity.  One must consequently impose
boundary conditions at infinity to formulate a sensible field theory.

This problem has been analyzed carefully in 3+1 dimensions by Avis,
Isham, and Storey \cite{Avis}, who show that there are three reasonable
boundary conditions for a scalar field at spatial infinity: Dirichlet
(D), Neumann (N), and ``transparent'' (T) boundary conditions.  The
latter---essentially a linear combination of Dirichlet and Neumann
conditions---are most easily obtained by viewing \ads\  as half of an
Einstein static universe; physically they correspond to a particular
recirculation of momentum and angular momentum at spatial infinity.  The
same choices exist in 2+1 dimensions \cite{LifOr}.  In particular, for
a massless, conformally coupled scalar field, described by an action
\beq
I = -\int\!d^3x\,\sqrt{-g} \left(
  {1\over2}g^{\mu\nu}\partial_\mu\phi\partial_\nu\phi
  + {1\over16}R\phi^2 \right) ,
\label{e4}
\eeq
the adS Greens functions are
\begin{eqnarray}
G_A^{(D)} &=& {1\over4\pi\sqrt{2}}
  \left\{ \sigma^{-1/2} - [\sigma + 2\ell^2]^{-1/2} \right\} \nonumber\\
G_A^{(N)} &=& {1\over4\pi\sqrt{2}}
  \left\{ \sigma^{-1/2} + [\sigma + 2\ell^2]^{-1/2} \right\} \\
G_A^{(T)} &=& {1\over4\pi\sqrt{2}}\, \sigma^{-1/2} .\nonumber
\label{e2}
\end{eqnarray}
Here $\sigma(x,x')$ is the square of the distance between $x$ and $x'$
in the embedding space described in section \ref{GG}; in the coordinates
of eqn.~\rref{b3},
\beq
\sigma(x,x') = {1\over2}\left[
  (X_1-X_1')^2 - (T_1-T_1')^2 + (X_2-X_2')^2 - (T_2-T_2')^2 \right] .
\label{e3}
\eeq
A more general Greens function with ``mixed'' boundary conditions,
\beq
G_A^{(\alpha)} = {1\over4\pi\sqrt{2}}
  \left\{ \sigma^{-1/2} - \alpha[\sigma + 2\ell^2]^{-1/2} \right\} ,
\label{e2a}
\eeq
may also be considered \cite{ShirMaka}.  The corresponding BTZ Greens
functions are then obtained from the sum \rref{e1}.

For the static ($J\!=\!0$) BTZ black hole, Lifschytz and Ortiz \cite{LifOr}
and Shiraishi and Maki \cite{ShirMaka} have analyzed the quantum behavior
of a massless, conformally coupled scalar field, using the full range of
boundary conditions discussed above.  For each of these boundary conditions,
the Greens function obtained from the sum \rref{e1} can be shown to be
periodic in imaginary time, with a period $\beta$ given by
\beq
\beta^{-1} = (N^\perp)^{-1} T_0 , \qquad T_0 = {r_+\over2\pi\ell^2}
  = {\kappa\over2\pi} ,
\label{e5}
\eeq
where $\kappa$ is the surface gravity \rref{a12}.  As usual, $\beta$
may be interpreted as a local inverse temperature, corresponding to
a temperature $T_0$ of the black hole corrected by a red shift
factor $(g_{00})^{-1/2}$.  The relationship of $T_0$ and $\kappa$
is then exactly the same as in 3+1 dimensions.\footnote{Reznik had
already found the corresponding result for (2+1)-dimensional gravity
with a positive cosmological constant in 1992 \cite{Reznik}.}  By
studying the analyticity properties of the Greens functions \rref{e2},
Lifschytz and Ortiz argue that the relevant vacuum state is the
Hartle-Hawking vacuum.  Similar computations of the propagator and
the expectation value $\langle\phi^2\rangle$ for transparent boundary
conditions have also been performed in Ref.~\citen{ShirMak}.

For the action \rref{e4}, the stress-energy tensor is
\beq
T_{\mu\nu} = {3\over4}\nabla_\mu\phi\nabla_\nu\phi
  -{1\over4}g_{\mu\nu}g^{\rho\sigma}\nabla_\rho\phi\nabla_\sigma\phi
  -{1\over4}\phi\nabla_\mu\nabla_\nu\phi
  +{1\over4}g_{\mu\nu}g^{\rho\sigma}\phi\nabla_\rho\nabla_\sigma\phi
  -{1\over16\ell^2}g_{\mu\nu}\phi^2 ,
\label{e6}
\eeq
and the expectation value $\langle0|T_{\mu\nu}|0\rangle$ can be obtained
by differentiating a two-point function and taking coincidence limits:
for example, the expectation value of the first term in \rref{e6} is
$$
\lim_{x\rightarrow x'} {3\over4}\nabla_\mu\nabla_{\nu'} G(x,x') .
$$
Only the $n\!=\!0$ term in the sum \rref{e1} diverges as $x\!\rightarrow
\!x'$, and the stress-energy tensor may be renormalized by subtracting
off this term, in effect setting the energy in a pure anti-de Sitter
universe to  zero.  The resulting expectation values are discussed
in Ref.~\citen{LifOr} (for $J\!=\!0$, Neumann and Dirichlet boundary
conditions, and untwisted fields), Ref.~\citen{ShirMaka} (for $J\!=\!0$,
arbitrary mixed boundary conditions, and twisted and untwisted fields),
and Ref.~\citen{Steif} (for arbitrary $J$, transparent boundary conditions,
and untwisted fields).  For the static black hole, $\langle0|T_{\mu\nu}|
0\rangle$ is regular on the horizon $r\!=\!r_+$, but diverges as $1/r^3$
at $r\!=\!0$, indicating a breakdown of the semiclassical approximation
and the possible emergence of a genuine curvature singularity.  The
back reaction on the metric may be estimated away from $r\!=\!0$
\cite{LifOr,ShirMaka}; its effect is to increase the horizon size and
to slightly change the radial acceleration felt by a test particle, but
for large black hole masses the corrections are exponentially suppressed.

For the rotating black hole, an interesting new phenomenon arises at the
inner horizon \cite{Steif}.  The expectation value of the stress-energy
tensor (at least for transparent boundary conditions) now involves terms
of the form
$$\sum {a_n\over |d_n|^{5/2}}$$
where
\beq
d_n = \sigma(x,\Lambda^nx) .
\label{e7}
\eeq
As a consequence, $\langle0|T_{\mu\nu}|0\rangle$ blows up on an infinite
sequence of timelike ``polarized hypersurfaces'' composed of points
$x$ connected to their images $\Lambda^nx$ by null geodesics.  In the
limit $n\!\rightarrow\!\infty$, these surfaces approach the inner horizon
$r=r_-$.  The back reaction of the quantum stress-energy tensor on the
metric may now be estimated; one finds that metric perturbations at a
geodesic distance $s$ from a polarized hypersurface grow as $s^{-1/2}$.
This result, which is strongly reminiscent of both the classical
instability of section \ref{Int} and of ``chronology protection''
in (3+1)-dimensional models \cite{Hawking}, strongly suggests that the
inner horizon is quantum mechanically unstable.

We may obtain further information about scalar fields in a black hole
background by computing the Bogoliubov coefficients between anti-de Sitter
``in'' states in the far past and BTZ ``out'' states.  The relevant
transformations have been analyzed by Hyun, Lee, and Yee \cite{Hyunb}
for the rotating black hole with transparent boundary conditions.  A
calculation of the expectation value of the ``out'' number operator in
the ``in'' vacuum state yields
\beq
{}_{\hbox{\scriptsize in}}\langle0|
N_{\omega m}^{\hbox{\scriptsize out}}|0\rangle_{\hbox{\scriptsize in}}
= {1\over \exp\left[2\pi(\omega-m\Omega_H)\beta_0\right] - 1} ,
\label{e8}
\eeq
where $\Omega_H=-N^\phi(r_+)$ is the angular velocity of the
horizon\footnote{The $\omega$ and $m$ dependence in \rref{e8} reflects
the structure \rref{a11} of the Killing vector normal to the horizon.}
and
\beq
\beta_0^{-1} = {r_+^2-r_-^2\over2\pi r_+\ell^2} = {\kappa\over2\pi} ,
\label{e9}
\eeq
reducing to \rref{e5} when $J=0$.  The thermal form of \rref{e8} is
an indication of Hawking radiation, while the dependence on $\omega-
m\Omega_H$ rather than $\omega$ is a sign that the rotating BTZ black
hole, like the Kerr black hole, has super-radiant modes.

The quantum field theory of a massless fermion in a static BTZ black
hole background has been studied as well \cite{Hyun}.  Hyun, Song, and
Lee compute the two-point function for Dirichlet and Neumann boundary
conditions at spatial infinity, again using the method of images
\rref{e1}, and study the response of a particle detector.  As in the
case of the conformally coupled massless scalar, the response is thermal,
with a temperature given by \rref{e5}.  Curiously, the response function
for fermions exhibits a Planck distribution, while the response function
for the scalar field exhibits a Fermi-Dirac distribution.  This
``statistics flip'' does not occur in the expectation value $\langle
N\rangle$ of the number operator \cite{Hyunb}, and the phenomenon is not
understood.

In a recent preprint \cite{Ichi}, Ichinose and Satoh have also studied
massive, non-conformally coupled scalar fields in a BTZ background.
They show that the two-point functions for the Hartle-Hawking vacuum
are again thermal, with a periodicity in imaginary time given by
\rref{e9}.  They also compute thermodynamic quantities such as the free
energy and entropy of the scalar field, but find that the results are
generally divergent and regularization-dependent.

\section{Thermodynamics \label{Therm}}
\setcounter{footnote}{0}

The results of the last section give a strong indication that the BTZ
black hole, like the Kerr black hole, is a thermodynamic object with
a temperature $T_0\!=\!\kappa/2\pi$.  In particular, the appearance of
thermal Greens functions and the particle distribution \rref{e8} both
point to the presence of thermal Hawking radiation.  These results rely,
however, on an unphysical splitting of the system into a classical
gravitational background and quantum matter fields.  In this section,
I will briefly review four approaches to black hole thermodynamics that
depend on the gravitational configuration alone: the Euclidean path
integral, the microcanonical ensemble of Brown and York, the method
of Noether charges, and the quantum tunneling approach.  (For a
comparison of several approaches to black hole entropy in a general
setting, see \cite{IWald2}.)

The most straightforward approach to black hole thermodynamics is the
Euclidean path integral of Gibbons and Hawking \cite{GibHawk}, in which
the black hole partition function is expressed as a path integral
periodic in imaginary time.  In 3+1 dimensions, it is well known that
the Euclidean black hole solution exists globally only when a particular
periodicity is imposed, and that this periodicity determines a unique
inverse temperature $\beta_0$.  To see that the same is true in 2+1
dimensions, we must study the Riemannian continuation of the BTZ metric
\rref{a1}, obtained by letting $t\!=\!i\tau$ and $J\!=\!iJ_E$:
\beq
ds_E^2 = (N_E^\perp)^2 d\tau^2 + f_E^{-2}dr^2
 + r^2\left( d\phi + N_E^\phi d\tau\right)^2
\label{f1}
\eeq
with
\beq
N_E^\perp = f_E
  = \left( -M + {r^2\over\ell^2} - {J_E^2\over4r^2}\right)^{1/2} ,
\qquad N_E^\phi = -iN^\phi = - {J_E\over2r^2} \, .
\label{f2}
\eeq
The Euclidean lapse function now has roots
\begin{eqnarray}
r_+ &=& \left\{{M\ell^2\over2}\left[ 1 +
 \left(1+{J_E^2\over M^2\ell^2}\right)^{1/2} \right]\right\}^{1/2},
\nonumber\\
r_- &=& -i|r_-| = \left\{{M\ell^2\over2}\left[ 1 -
 \left(1+{J_E^2\over M^2\ell^2}\right)^{1/2} \right]\right\}^{1/2} .
\label{f3}
\end{eqnarray}
Note that the continuation of $J$ to imaginary values, necessary for
the metric \rref{f1} to be Riemannian, is physically sensible, since
the angular velocity is now a rate of change of a real angle with
respect to imaginary time.

The metric \rref{f1}--\rref{f2} is a positive definite metric of
constant negative curvature, and the spacetime is therefore locally
isometric to hyperbolic three-space $\IH^3$.  This isometry is most
easily exhibited by means of the Euclidean analogue of the coordinate
transformation \rref{b10}, which is now globally valid \cite{CarTeit},
\begin{eqnarray}
x &=& \left({r^2-r_+^2\over r^2-r_-^2}\right)^{1/2}
      \cos\left( {r_+\over\ell^2}\tau + {|r_-|\over\ell}\phi \right)
      \exp\left\{ {r_+\over\ell}\phi - {|r_-|\over\ell^2}\tau \right\}
      \nonumber \\
y &=& \left({r^2-r_+^2\over r^2-r_-^2}\right)^{1/2}
      \sin\left( {r_+\over\ell^2}\tau + {|r_-|\over\ell}\phi \right)
      \exp\left\{ {r_+\over\ell}\phi - {|r_-|\over\ell^2}\tau \right\} \\
z &=& \left({r_+^2-r_-^2\over r^2-r_-^2}\right)^{1/2}
      \exp\left\{ {r_+\over\ell}\phi - {|r_-|\over\ell^2}\tau \right\} .
      \nonumber
\label{f4}
\end{eqnarray}
In these coordinates, the metric becomes that of the standard upper
half-space representation of hyperbolic three-space,
\beq
ds_E^2 = {\ell^2\over z^2}(dx^2+dy^2+dz^2)
  = {\ell^2\over\sin^2\chi}\left(
  {dR^2\over R^2} + d\chi^2 + \cos^2\chi d\theta^2\right) ,\qquad (z>0).
\label{f5}
\eeq
Here $(R,\theta,\chi)$ are standard spherical coordinates for the upper
half-space $z\!>\!0$,
\beq
(x,y,z) = (R\cos\theta\cos\chi,R\sin\theta\cos\chi,R\sin\chi) ,
\label{f5a}
\eeq
and periodicity in the Schwarzschild angular coordinate $\phi$ requires
that we identify
\beq
(R,\theta,\chi) \sim
  (Re^{2\pi r_+/\ell}, \theta + {2\pi|r_-|\over\ell}, \chi) .
\label{f6}
\eeq
Eqn.~\rref{f6} is the Euclidean version of the identifications $\langle
(\rho_R,\rho_L)\rangle$ of section \ref{GG}.  A fundamental region is the
space between two hemispheres $R\!=\!1$ and $R\!=\!e^{2\pi r_+/\ell}$,
with the inside and outside boundaries identified by a translation along
a radial line followed by a $2\pi|r_-|/\ell$ rotation around the $z$ axis
(see figure 3).  The topology is thus $\IR^2\!\times\!S^1$, as expected.

Now, for the coordinate transformation \rref{f4} to be nonsingular at
the $z$ axis $r=r_+$, we must require periodicity in the arguments of
the trigonometric functions; that is, we must identify
\beq
(\phi,\tau) \sim (\phi+\Phi,\tau+\beta_0)
\label{f7}
\eeq
where
\beq
\beta_0 = {2\pi r_+\ell^2\over r_+^2 - r_-^2} ,\qquad
\Phi =  {2\pi |r_-|\ell\over r_+^2 - r_-^2} \, .
\label{f8}
\eeq
Equivalently, if we define a shifted angular coordinate $\phi'\!=\!
\phi-(\Phi/\beta_0)\tau$, the identifications become $(\phi',\tau)\!
\sim\!(\phi', \tau+\beta_0)$.  We thus obtain exactly the same temperature
\rref{e9} that appeared in the analysis of quantum fields.  If this
requirement of periodicity is lifted, the Euclidean black hole acquires
a conical singularity at the horizon, whose nature I will discuss further
in section \ref{SM}.

To obtain the entropy from the Euclidean path integral, we must now
evaluate the grand canonical partition function
\beq
Z = \int[dg]\,e^{I_E[g]} ,
\label{f9}
\eeq
where $I_E$ is the Euclidean action.  Normally, only the classical
approximation $Z\!\sim\!\exp\{I_E[\bar g]\}$ is considered, where $\bar g$
is the extremal metric \rref{f1}.  The action $I$ includes boundary
terms at $r=r_+$ and $r=\infty$, which are analyzed carefully in
Ref.~\citen{CarTeit} for the (2+1)-dimensional black hole (see also
\cite{BTZcont} for a more general discussion).  For a temperature
$\beta_0$ and a rotational chemical potential $\Omega$, one finds an
action
\beq
I_E[\bar g] = 4\pi r_+ - \beta_0 (M - \Omega J) ,
\label{f10}
\eeq
corresponding to an entropy $4\pi r_+$, or with factors of $\hbar$ and
$G$ restored,
\beq
S = {2\pi r_+\over4\hbar G} .
\label{f11}
\eeq
Note the similarity with the standard Bekenstein entropy, which like
\rref{f11} is one-fourth of the horizon size in Planck units.

In contrast to the (3+1)-dimensional theory, one can also compute
the first quantum correction to the path integral \rref{f9}---the Van
Vleck-Morette determinant---by taking advantage of the Chern-Simons
formalism.  This calculation was performed, in a different context, in
Ref.~\citen{Carlip2}; to the next order, the partition function becomes
\cite{CarTeit}
\beq
Z \sim \exp\left\{ I[\bar g] + {2\pi r_+\over\ell} \right\},
\label{f12}
\eeq
leading naively to an entropy
\beq
S = {2\pi r_+\over4\hbar G}\left( 1 + {4\hbar G\over\ell}\right) .
\label{f13}
\eeq
Ghosh and Mitra have argued that this expression does not properly account
for quantum corrections to $\beta_0$, and that the true one-loop-corrected
temperature and entropy should be \cite{Ghosh}
\beq
\beta_0 = {1\over8\hbar G}{2\pi r_+\ell^2\over r_+^2 - r_-^2}
  \left( 1 + {8\hbar G\over\ell}\right) , \qquad
S = {2\pi r_+\over4\hbar G}\left( 1 + {8\hbar G\over\ell}\right) .
\label{f14}
\eeq

The Euclidean path integral offers a simple and attractive approach
to black hole thermodynamics, but it has important limitations.  In
particular, the canonical and grand canonical ensembles do not always
exist for arbitrarily large gravitating systems.  Brown and York have
recently developed a more sophisticated path integral approach
\cite{BYork2}, based on the microcanonical ensemble for a black hole
in a cavity.  The resulting microcanonical partition function can again
be written in the form \rref{f9}, where the action $I$ has the same
``bulk'' terms as in the Gibbons-Hawking approach, but different boundary
terms.\footnote{A fundamental feature of general relativity is that
the energy and angular momentum in a finite spatial region $U$ may be
expressed in terms of integrals over the boundary $\partial U$; this is
the underlying reason that the microcanonical ensemble can be obtained
from the canonical ensemble by adjusting only the boundary terms in
the path integral.}

This approach to the BTZ black hole is discussed in detail by Brown,
Creighton, and Mann in Ref.~\citen{BMann}.  They find that the temperature
at the boundary of a cavity of radius $R$ is
\beq
T = {1\over N^\perp(R)}{\kappa\over2\pi} ,
\label{f15}
\eeq
the correct red-shifted form of the temperature \rref{e9}, and that
the thermodynamic chemical potential conjugate to $J$ is
\beq
\Omega = {1\over N^\perp(R)}\left( N^\phi(R)-N^\phi(r_+)\right) ,
\label{f16}
\eeq
in agreement with the results of \cite{CarTeit}.  Moreover, with the
entropy \rref{f11}, the (2+1)-dimensional black hole obeys the standard
first law of thermodynamics,
\beq
dE = TdS + \Omega dJ - {\cal P}d(2\pi R) ,
\label{f17}
\eeq
where $\cal P$ is the surface pressure at $R$.

A third approach to black hole thermodynamics is that of Wald
\cite{Wald0}, who defines the entropy as the Noether charge associated
with the Killing vector $\xi$ that is normal to a bifurcate event
horizon and is normalized to unit surface gravity.  The argument
for this definition, based on the application of the first law of
thermodynamics to diffeomorphism-invariant systems, is rather long,
and I will not try to reproduce it here.  For (2+1)-dimensional
gravity, the relevant Noether charge one-form is (see \cite{IWald3})
\beq
Q = -{1\over16\pi G} \epsilon_{abc}\nabla^b\xi^cdx^a ,
\label{f18}
\eeq
where the Killing vector $\xi$ is determined from \rref{a11} by $\xi^a\!
=\!\kappa^{-1}\chi^a$.  The one-form \rref{f18} is to be integrated over
the bifurcation circle $r\!=\!r_+$, with the volume element induced by
the metric \rref{a9}.  It is straightforward to check that up to slight
differences in normalization, the entropy \rref{f11} is again reproduced.
The same computation has been carried out in the first-order formalism
in Ref.~\citen{CGeg}.

Finally, let me briefly mention a fourth approach to black hole
thermodynamics, developed by Casher and Englert \cite{CashEng}.
It is well known that the Wheeler-DeWitt equation in quantum gravity
has no preferred time parameter, but that in a region in which the
gravitational field is nearly classical, the WKB approximation leads
to an effective ``time''  for matter determined by the classical
gravitational trajectories \cite{WKB}.  Casher and Englert argue that
the thermodynamic properties of black holes reflect the breakdown of
this approximation, and the consequent failure of semiclassical
unitarity, in regions of superspace in which tunneling processes
are important.  Applying this argument to the BTZ black hole, Englert
and Reznik find that the relevant tunneling process in 2+1 dimensions
is one between a thick spherical shell of matter (held in equilibrium
by a uniform pressure $|p_r|\!=\!|p_\theta|\!=\!\rho$) and a black hole
of the same mass \cite{EngRez}.  They compute a tunneling entropy and
find an expression that is once again equal to that of eqn.~\rref{f11}.

These various results may now be used to determine thermodynamic
quantities such as heat capacities \cite{BMann,Zas}.  The microcanonical
approach is especially useful for this purpose.  In contrast to the
(3+1)-dimensional black hole, it may be shown that the BTZ black hole
has a strictly positive heat capacity $C_{R,J}$ at fixed $R$ and $J$,
and that the temperature $T$ is a monotonically increasing function of
$r_+$.  Consequently, there is a unique solution $M(\beta,R,J)$ for a
black hole in a cavity at fixed surface temperature and angular momentum,
and the resulting black hole is thermodynamically stable---there are no
negative-heat-capacity instantons like those that occur in 3+1 dimensions.
The grand canonical ensemble ($\Omega$ and $\beta$ fixed) is similarly
stable.  On the other hand, pressure ensembles ($\cal P$ fixed, $R$
allowed to vary) are thermodynamically unstable, with negative heat
capacities \cite{Zas}.  One may also analyze the thermodynamics of a
gas of noninteracting (2+1)-dimensional black holes; if one interprets
the entropy \rref{f11} as the logarithm of the number of states of a
single black hole, Cai et al.\ have shown that it is possible to construct
microcanonical and canonical partition functions for such a system
\cite{Cai}.

A natural question, of fundamental importance in 3+1 dimensions, is
whether the black hole evaporates completely, and if not, what final
states are possible.  Here, unfortunately, the (2+1)-dimensional model
may not be too helpful.  Note first that the temperature of a BTZ black
hole is $T\!\sim\!M^{1/2}$, which, in contrast to the (3+1)-dimensional
case, goes to zero as $M$ decreases.  Naively, complete evaporation
would take an infinite amount of time; indeed, the energy flux in
thermal radiation in 2+1 dimensions is proportional to the cube of
the temperature, and Stefan's law gives \cite{HorWelch}
\beq
{dM\over dt} \sim M^2 ,
\label{f19}
\eeq
or $M\!\sim\!1/t$.  Moreover, the unusual boundary conditions for
radiation in an asymptotically anti-de Sitter space, discussed in
section \ref{QFT}, affect the thermodynamics; Reznik has argued that
for a space with such asymptotic behavior, a black hole with a large
enough initial mass will come to equilibrium with thermal radiation
even in an infinitely large universe \cite{Rez2}.  Further work on this
problem---including, ideally, a detailed semiclassical computation of
the interaction of a BTZ black hole with Hawking radiation with various
boundary conditions at infinity---may be feasible, and would be of great
interest.

\section{Quantization and Statistical Mechanics \label{SM}}
\setcounter{footnote}{0}

Several of the derivations of black hole entropy described in the
preceding section were quantum mechanical, in the sense that they were
based on the (Euclidean) path integral.  They made only limited use of
the machinery of quantum mechanics, however, relying instead on
classical (and, in one case, one-loop) approximations.  It is natural to
ask whether a full quantum mechanical treatment of the (2+1)-dimensional
black hole can provide us with further insight.  In particular, we might
hope that such a treatment could give us a ``statistical mechanical''
explanation of black hole entropy in terms of the microscopic physics
of quantum gravitational states.  We are still far from having a complete
answer to this question, but several interesting first steps have been
taken.

Perhaps the simplest starting point is a minisuperspace quantization of
metrics of the form \rref{a1} or \rref{f1}, with $N^\perp$, $f$, and
$N^\phi$ treated as arbitrary functions of $r$.  This model has been
analyzed in Ref.~\citen{CarTeit} in the Euclidean setting.  The
natural approach to the quantum theory is that of radial quantization,
in which the radial coordinate $r$ is used as a ``time'' variable
and canonical commutation relations are imposed on surfaces of
constant $r$.  In terms of the parameters of \rref{f1}, one finds
two pairs of conjugate variables, $(\beta,f_E^2)$ and $(N^\phi,p)$,
where
\beq
\beta(r) = f_E^{-1}(r)N^\perp_E(r)
\label{g1}
\eeq
and $p(r)$ is the $r$-$\phi$ component of the gravitational momentum.
The quantum theory may be obtained by imposing equal $r$ commutation
relations at $r\!=\!r_+$,
\beq
[\beta(r_+),f^2(r_+)] = [N^\phi(r_+),p(r_+)] = i\hbar ,
\label{g2}
\eeq
and describing the evolution by a ``time''-dependent radial Hamiltonian
that may be shown to have the form
\beq
H_r(r) = -\beta(r)\left[ {p^2(r)\over2r^3} + {2r\over\ell^2}\right] .
\label{g3}
\eeq
In principle, we should now be able to compute the partition function
\rref{f9} by taking the trace of the propagator $K[{}^{(2)}{\cal G}_2,
{}^{(2)}{\cal G}_1;r_+,p_+;\beta(\infty),\tilde N^{\phi'}(\infty)]$
in this quantum theory.  The exact calculation has not been performed,
but it is argued in \cite{CarTeit} that the classical approximation
duplicates the result found in section \ref{Therm} for the entropy.

To understand the canonical variables in this minisuperspace, it is
useful to turn to the Chern-Simons description of (2+1)-dimensional
gravity, as described in section \ref{GG}.  For a Euclidean black hole,
the holonomy \rref{b9} may be analytically continued to become an
$\hbox{SL}(2,\IC)$ holonomy
\beq
\rho[\gamma] = \left( \begin{array}{cc}
 e^{\pi (r_+ + i|r_-|)/\ell} & 0 \\ 0 & e^{-\pi (r_+ + i|r_-|)/\ell}
 \end{array}\right) ,
\label{g4}
\eeq
where $\gamma$ is the closed curve \rref{b16a}; in figure 3, such a
curve is represented by the portion of the line segment $L$ lying
between the two (identified) hemispheres.  At first sight, there appears
to be no other closed curve around which to define a holonomy; the
topology of the Euclidean black hole is $\IR^2\!\times\!S^1$, a manifold
whose fundamental group has only a single generator.  Were this the case,
the Chern-Simons quantum theory would be trivial: $r_+$ and $r_-$ could
be treated as operators, but they would have no canonical conjugates.

For the computation of transition amplitudes and propagators, however,
the relevant spacetime is not the complete Euclidean black hole, but
rather the wedge $\tau_1\!\le\!\tau\!\le\!\tau_2$, whose geometry in
the upper half-space representation is depicted in figure 4.  We can
now consider a line segment $\delta$ connecting the points $(R_1,
\theta_1)$ and $(R_2\!=\!e^\Sigma R_1, \theta_2\!=\!\theta_1+ \Theta)$
along a surface of constant ``radius'' $\chi$.  (It may be seen from
\rref{f4} and \rref{f5a} that constant $\chi$ does in fact correspond
to constant radius $r$.)  Such a curve is exemplified in figure 4 by
the portion of the line segment $K$ lying between the edges $\tau\!=\!
\tau_1$ and $\tau\!=\!\tau_2$.  A simple computation then gives an
$\hbox{SL}(2,\IC)$ holonomy
\beq
\rho[\delta] = \left( \begin{array}{cc}
 e^{\pi(\Sigma + i\Theta)/\ell} & 0 \\ 0 & e^{-\pi(\Sigma + i\Theta)/\ell}
 \end{array}\right) .
\label{g5}
\eeq
For the complete spacetime of figure 3, values of $\Sigma$ and $\Theta$
other than $\Sigma\!=\!0$, $\Theta\!=\!2\pi$ signal the presence of a
conical singularity with a ``helical twist'' at $\chi\!=\!\pi/2$, that
is, at the horizon $r\!=\!r_+$.

The minisuperspace variables $f_E$, $p$, $\beta$, and $N^\phi$ may now
be written in terms of the constants of motion $r_\pm$, $\Theta$, and
$\Sigma$ that describe the holonomies (see Ref.~\citen{CarTeit} for the
complete expressions).  The commutators \rref{g2} then induce commutators
\begin{eqnarray}
\left[ r_+, \Theta \right] &=& \left[ |r_-|, \Sigma \right]
  = 4i\hbar G \nonumber\\
\left[ |r_-|, \Theta \right] &=& \left[ r_+, \Sigma \right] = 0 ,
\label{g6}
\end{eqnarray}
providing a starting point for quantization of the Chern-Simons theory.
The same commutators may be obtained directly from the holonomies,
using the results of Nelson and Regge for quantization of holonomies in
(2+1)-dimensional gravity \cite{NR}.  Observe in particular that the
horizon radius $r_+$ and the deficit angle $\Theta$ at the horizon are
conjugate operators, implying that it is inconsistent to simply set
$\Theta$ to $2\pi$ in the quantum theory.  This role of the horizon
deficit angle as a canonical variable is the starting point for the
approach of Ref.~\citen{CarTeit2} to an extended Wheeler-DeWitt
equation for black holes in arbitrary dimensions.

Note that for a Chern-Simons theory---at least on a manifold without
boundary---the holonomies of the connection provide a complete set of
physical observables.  For the Euclidean BTZ black hole topology, this
implies that the observables $(r_+,\Theta)$ and $(r_-,\Sigma)$, which
parametrize the geometry of the horizon, should be sufficient for a
complete quantum theory; the Chern-Simons formulation is {\em not\/}
a minisuperspace model.  In particular, if the entropy of the black hole
can be explained in terms of microscopic quantum gravitational states,
the relevant physics should involve states at the horizon depending on
these canonical pairs of variables alone.

It is possible, however, that this conclusion is an artifact of
Euclideanization, since the continuation of the BTZ metric to Riemannian
signature causes the two-dimensional Lorentzian horizon to collapse to
a one-dimensional circle, ``compressing'' the horizon dynamics.  To see
whether this is a problem, we must investigate the dynamics of the
horizon of the Lorentzian black hole.  Here, the Chern-Simons formalism
proves to be a powerful tool.

Consider the gravitational action \rref{b14} on a spacetime consisting
of the exterior of a (2+1)-dimensional black hole, with the event horizon
acting as a boundary $\partial M$.  It is well known that a Chern-Simons
theory on a manifold with boundary is no longer a theory of a finite
number of holonomies alone; rather, the Chern-Simons action induces
a two-dimensional Wess-Zumino-Witten (WZW) action on $\partial M$
\cite{MS,EMSS}.  This phenomenon occurs because the presence of a
boundary partially breaks the gauge symmetry, allowing would-be ``pure
gauge'' degrees of freedom to become dynamical.

For the case of (2+1)-dimensional gravity with a negative cosmological
constant, this phenomenon is discussed in Ref.~\citen{Carstat}.  The
requirement that $\partial M$ be an apparent horizon imposes boundary
conditions
\beq
A^+_\phi = A^+_v = \tilde A^+_\phi = \tilde A^+_v = 0, \quad
A^2_\phi = \bar\omega + {r_+\over\sqrt{2}\ell} , \quad
\tilde A^2_\phi = \bar\omega - {r_+\over\sqrt{2}\ell}
\label{g7}
\eeq
on the gauge fields \rref{b13}.  Here $r_+$ is the horizon radius,
while the interpretation of $\bar\omega$ is less clear; it is argued
in Ref.~\citen{Carstat} that one should sum over constant values of
$\bar\omega$ to count macroscopically indistinguishable states.  With
these boundary conditions, the induced action on $\partial M$ may be
shown to take the form (up to possible finite renormalizations)
\beq
I[g,\tilde g] =
 - kI^{\raise2pt\hbox{$\scriptstyle +$}}_{\hbox{\scriptsize WZW}}[g,A]
 + kI^{\raise2pt\hbox{$\scriptstyle +$}}_{\hbox{\scriptsize WZW}}
 [\tilde g,\tilde A] ,
\label{g8}
\eeq
where $I^{\raise2pt\hbox{$\scriptstyle +$}}_{\hbox{\scriptsize WZW}}[g,A]$
is the $\hbox{SO}(2,1)$ chiral Wess-Zumino-Witten action,
\begin{eqnarray}
I^{\raise2pt\hbox{$\scriptstyle +$}}_{\hbox{\scriptsize WZW}}[g,A]
  &=& {1\over4\pi}\int_{\partial M}\hbox{Tr}
  \left(g^{-1}\partial_\phi g\right)\left(g^{-1}\partial_v g\right)\
  \nonumber\\
  &\ & +\,{1\over2\pi}\int_{\partial M}\hbox{Tr}
  \left(g^{-1}\partial_v g\right)\left(g^{-1} A_\phi g\right)
  + {1\over12\pi}\int_M\hbox{Tr}\left(g^{-1}dg\right)^{\lower2pt
  \hbox{$\scriptstyle 3$}}
\label{g9}
\end{eqnarray}
and the constant $k$ is
\beq
k = {\ell\sqrt{2}\over 8G} \, .
\label{g10}
\eeq
The effective quantum theory of the horizon is thus described by a
pair of $\hbox{SO}(2,1)$ WZW actions, and one may attempt to find the
entropy by counting the resulting states.

Such $\hbox{SO}(2,1)$ WZW models are not yet completely understood.
In the limit of large $k$ (or small $\Lambda$), however, the action
\rref{g8} may be approximated by a system of six independent bosonic
string oscillators.  Such a system has an infinite number of states,
but most of these are eliminated by a remaining gauge symmetry---a
remnant of the Wheeler-DeWitt equation---that expresses invariance
under time-dependent shifts of the angular coordinate $\phi$.  These
transformations are generated by the Virasoro operator $L_0$.  Acting
on a naturally defined Hilbert space (built by a ``highest weight''
construction), $L_0$ takes the form \cite{Carstat}
\beq
L_0 = N
  + {4k^2\over4k^2-1}\left(
    \bar\omega - {\sqrt{2}kr_+\over\ell}\right)^2
  - {2k^2r_+^2\over\ell^2} ,
\label{g11}
\eeq
where $N\!=\!\sum_{i=1}^6 N_i$ is a number operator for the six
``stringy'' oscillators.  Requiring that $L_0$ annihilate physical
states thus determines $N$ in terms of $r_+$.

It is now a standard result of number theory \cite{Ram} and string
theory \cite{Wein} that the number of states of such a system behaves
asymptotically as
\beq
n(N) \sim \exp\left\{\pi\sqrt{6\cdot{2N\over3}}\right\} .
\label{g12}
\eeq
Using (\ref{g11}) to determine $N$, integrating over $\bar\omega$, and
inserting (\ref{g10}) for $k$, we obtain
\beq
\log n(r_+) \sim {2\pi r_+\over 4\hbar G} ,
\label{g13}
\eeq
which is precisely the expression for the entropy of the (2+1)-dimensional
black hole derived in the preceding section.

Unfortunately, this argument relies heavily on the particular
characteristics of general relativity in 2+1 dimensions, and does not
directly generalize to higher dimensions.  A possible guess for a
(3+1)-dimensional analogue of the action \rref{g8} would be an induced
action on the horizon for ``would-be diffeomorphisms'' generated by vector
fields with components normal to the horizon.  The resulting picture
is reminiscent of the membrane model of Maggiore \cite{Mag,Mag2} and
the ``quantized normal mode'' picture of York \cite{York}, but the
analogies are so far only suggestive.

Given this picture of black hole thermodynamics, we may now ask whether
it is possible to obtain further information about the quantum mechanics
of the horizon degrees of freedom.  Little is known about this question,
but there have been some interesting speculations.  Maggiore has recently
proposed an effective action for the horizon, using the position
$\zeta^\mu(x)$ of the horizon as a set of collective coordinates, and
has obtained a minisuperspace Schr\"odinger equation for spherical
fluctuations \cite{Mag2}.  By considering a related string model, Kogan
has argued that the quantum black hole should have a discrete mass
spectrum, with $r_+^2$ quantized in integral multiples of the Planck
length \cite{Kogan}.

\section{Generalizations \label{Ass}}
\setcounter{footnote}{0}

The focus of this paper has been on the simplest form of the
(2+1)-dimensional black hole, that of Ba\~nados, Teitelboim, and Zanelli.
If one introduces additional fields, however, a number of interesting
generalizations become possible.  In this section, I will briefly review
a few of these generalized black holes: electrically charged black holes,
dilatonic black holes, black holes arising in string theory, black holes
in topologically massive gravity, and black holes formed from
``topological'' matter.

The simplest extension of the BTZ black hole may be found by
coupling an electromagnetic field to obtain a (2+1)-dimensional
``Reissner-N{\"o}rdstrom'' solution.  For the static case ($N^\phi\!=\!0$),
we can take as the electromagnetic potential the one-form
\beq
A = -Q\ln(r/r_0) dt ,
\label{h1}
\eeq
and modify the metric \rref{a1}--\rref{a2} by setting
\beq
N^\perp = f
  = \left( -M + {r^2\over\ell^2} + {J^2\over4r^2}
  - {1\over2}Q^2\ln(r/r_0) \right)^{1/2}
\label{h2}
\eeq
to obtain a solution of the (2+1)-dimensional Einstein-Maxwell equations.
Unlike the ordinary BTZ black hole, this solution exists even when the
cosmological constant vanishes, reducing to the metric originally
discovered independently by Deser and Mazur \cite{DesMaz}, Gott, Simon,
and Alpert \cite{Gott}, and Melvin \cite{Melvin}.

Not surprisingly, this construction fails when $J\!\ne\!0$: a rotating
black hole should have a nonvanishing magnetic field, and the form
\rref{h1} of the vector potential must be generalized.  As far as I know,
the general solution for a charged rotating black hole in 2+1 dimensions
is not known.  However, the particular case of an extreme ($J\!=\!\pm
M\ell$) black hole with a mass $M\!=\!8\pi GQ^2$ and a self dual or
anti-self dual electromagnetic field ($E\!=\!\pm B$) has recently been
investigated by Kamata and Koikawa \cite{Kamata}.

The stress-energy tensor for the charged black hole is, of course,
nonzero, so by \rref{b1}, the spacetime is no longer one of constant
curvature.  In particular, it is not possible to express the charged
black hole as a quotient of anti-de Sitter space by a group of
isometries.  Nevertheless, the spacetime remains simple enough that
explicit quantum field theory calculations of the type described in
section \ref{QFT} may still be possible; work on this problem is in
progress \cite{Qayum}.

A further extension of the BTZ black hole may be obtained by introducing
a dilaton coupling.  Chan and Mann \cite{ChanMann2} have investigated
black hole solutions for an action of the form
\beq
I = \int\!d^3x\,\sqrt{-g}\left(R - {B\over2}\nabla_\mu\phi\nabla^\mu\phi
  - e^{-4a\phi}F_{\mu\nu}F^{\mu\nu} + 2e^{b\phi}\Lambda \right) ,
\label{h3}
\eeq
where $\phi$ is the dilaton field, $F_{\mu\nu}$ is the ordinary
electromagnetic field, and the coupling strengths $a$, $b$, and $B$
are arbitrary.  By adjusting the couplings, they find a one-parameter
family of static black holes with dilaton fields of the form $\phi\!=
\!k\ln(r/r_0)$, exhibiting a wide variety of horizon structures.  The
temperatures, quasilocal energies, and entropies of these solutions
are fairly easy to compute, and in contrast to the BTZ black hole---but
in analogy to the (3+1)-dimensional black hole---one can find solutions
with negative heat capacities.  An action equivalent to \rref{h3} with
$F_{\mu\nu}\!=\!0$ and $b\!=\!4$ has also been studied by S{\'a} et
al.\ \cite{Sa}, who examine the structure of horizons and geodesics
for a range of values of the coupling constant $B$.  (The action of
Ref.~\citen{Sa} differs from that of Chan and Mann by a rescaling
$g_{\mu\nu}\!\rightarrow\!e^{4\phi}g_{\mu\nu}$ of the metric.)

Perhaps the most interesting generalization of the BTZ solution comes
from its connection with string theory \cite{HorWelch,Kaloper}.  The
low energy string effective action is
\beq
I = \int\!d^3x\,\sqrt{-g}e^{-2\phi}\left( {4\over k} + R
  + 4\nabla_\mu\phi\nabla^\mu\phi
  - {1\over12}H_{\mu\nu\rho}H^{\mu\nu\rho} \right) ,
\label{h4}
\eeq
where $\phi$ is the dilaton and $H_{\mu\nu\rho}$ is an antisymmetric
Kalb-Ramond field, which in three dimensions must be proportional to the
volume form $\epsilon_{\mu\nu\rho}$.  Horowitz and Welch \cite{HorWelch}
point out that in three dimensions, the ansatz
\beq
H_{\mu\nu\rho} = {2\over\ell}\epsilon_{\mu\nu\rho}, \quad \phi=0 ,
\quad k=\ell^2
\label{h5}
\eeq
reduces the equations of motion coming from \rref{h4} to the Einstein
field equations \rref{a3}.  The BTZ black hole metric is thus a part of
a solution of low energy string theory.

In fact, there is a corresponding {\em exact\/} solution of string
theory.  As we saw in section \ref{GG}, the (2+1)-dimensional black hole
can be represented as a quotient of the group manifold $\hbox{SL}(2,\IR)$
by a discrete group of isomorphisms.  On the other hand, an $\hbox{SL}
(2,\IR)$ WZW model with an appropriately chosen central charge is an
exact string theory vacuum, describing the propagation of strings
on this same group manifold.  By quotienting out the discrete group
$\langle(\rho_L,\rho_R)\rangle$ of section \ref{GG} by means of an
orbifold construction, one obtains a theory that may be shown to be
an exact string theoretical representation of the BTZ black hole
\cite{HorWelch,Kaloper,Kal2}.

This stringy BTZ black hole provides an interesting model for studying
target space duality in string theory, that is, the existence of
physically equivalent string theories whose low energy limits may appear
to be highly inequivalent \cite{HorWelch,Ali,Hor2}.  The (2+1)-dimensional
asymptotically anti-de Sitter black hole is dual to an asymptotically
flat ``black string'' \cite{HorneHor} with an equal horizon circumference,
and it has been suggested that this may indicate that string theory is,
in some sense, insensitive to the presence of a cosmological constant
\cite{HorWelch}.  Similar duality transformations have been used to
construct a large family of three-dimensional stationary string solutions
with horizons \cite{AndKal}.  Ghoruku and Larsen have also studied tachyon
scattering in a stringy BTZ black hole background, reproducing the
standard Hawking temperature of section \ref{QFT} in a suitable limit
\cite{Gho}.

As yet another extension of the BTZ solution, one may consider black
holes in (2+1)-dimensional topologically massive gravity \cite{DJTemp},
that is, Einstein gravity with a gravitational Chern-Simons term,
\beq
I_{\hbox{\scriptsize GCS}}={k'\over4\pi}\int\!d^3x\,\epsilon^{\lambda\mu\nu}
  \Gamma^\rho_{\lambda\sigma}\left(\partial_\mu\Gamma^\sigma_{\rho\nu}
  + {2\over3}\Gamma^\sigma_{\mu\tau}\Gamma^\tau_{\nu\rho}\right) ,
\label{h6}
\eeq
added to the action.  Like the gauge Chern-Simons term \rref{b15}, the
action \rref{h6} is invariant---in this case under diffeomorphisms---even
though it depends explicitly on the connection.  The effect of such a
term is to reintroduce a local, propagating degree of freedom in
(2+1)-dimensional gravity.  As Kaloper points out \cite{Kaloper}, the
BTZ solution is also a solution of the field equations of topologically
massive gravity, since its Cotton tensor
\beq
C^{\mu\nu} = \epsilon^{\mu\rho\sigma}\nabla_\rho
  (R_\sigma^\nu - {1\over4}\delta_\sigma^\nu R)
\label{h7}
\eeq
vanishes identically.  Nutku has found an additional class of solutions
of topologically massive gravity with black-hole-like event horizons
\cite{Nutku}; these are not asymptotically anti-de Sitter, however,
and it is not clear that one can define such quantities as mass and
angular momentum.  A further class of solutions is discussed by
Cl{\'e}ment \cite{Clem2}.  These asymptotically approach extreme BTZ
black holes, but are geodesically complete and have no event horizons.
Their physical interpretation is unclear, but it would be interesting
to expore their possible role as stable end points of black hole
evaporation.

As a final generalization of the BTZ solution, one may consider black
holes in a model of (2+1)-dimensional gravity interacting with
``topological'' matter, that is, fields with finitely many degrees of
freedom that do not couple directly to the metric in the Lagrangian.
In Ref.~\citen{CGeg}, black holes in one such theory, with an action
\beq
I= \int_M\left(e^a\wedge R_a[\omega]+B^a\wedge D_\omega C_a\right) ,
\label{h8}
\eeq
are described.  Here $R_a$ is the curvature of the spin connection
$\omega$, and the first term is the standard first-order action for
Einstein gravity, but the cosmological constant has been replaced by a
pair of $\hbox{SO}(2,1)$-valued ``matter'' fields.  ($D_\omega$ is
the covariant exterior derivative.)  The configuration space of this
model is finite-dimensional---in fact, it is parametrized by a set of
$\hbox{ISO}(2,1)$-valued holonomies---and there are peculiar gauge
transformations that mix the gravitational and matter degrees of
freedom \cite{CarGeg}.  It is shown in Ref.~\citen{CGeg} that this
model admits a solution whose geometry is that of the BTZ black hole.
The mass, charge, and thermodynamic properties, computed by the method
of Noether charges, are rather strange, however, presumably because of
contributions of the $B$ and $C$ fields; the physics of the model is
not yet very well understood.

\section{Conclusion \label{Conc}}
\setcounter{footnote}{0}

The universe is not three-dimensional, and one must be cautious about
conclusions drawn from (2+1)-dimensional models.  Nevertheless, the
BTZ black hole is similar enough to the realistic Kerr solution that
its properties deserve to be taken seriously.  In the classical realm,
this model is perhaps most useful as a pedagogical tool---it allows us
to explore many of the general characteristics of black hole dynamics
in a framework in which we are not swamped by mathematical complications.
Thus, for example, we can investigate detailed models of collapsing
matter, mass inflation, and similar phenomena without having to resort
to numerical simulations.

It is in the quantum realm, however, that the power of the BTZ model
truly becomes evident.  In 3+1 dimensions, we simply do not have a
working theory of quantum gravity, and the study of black hole quantum
mechanics is necessarily approximate and speculative.  In 2+1 dimensions,
on the other hand, most of the obstructions to the quantization of general
relativity disappear, and we can hope to reach reliable conclusions about
quantum mechanical systems.  The differences between the (2+1)-dimensional
black hole and its (3+1)-dimensional counterpart---for example, the
positive specific heat of the BTZ solution---cannot be neglected, of
course, but the developing work on the BTZ black hole in quantum gravity
has the potential to have far-reaching impact.

\newpage
\begin{flushleft}
\large\bf Acknowledgements
\end{flushleft}

I would like to thank Alan Steif for a number of helpful suggestions.
This work was supported in part by National Science Foundation grant
PHY-93-57203 and Department of Energy grant DE-FG03-91ER40674.

\newpage
\begin{flushleft}
\large\bf Figure Captions
\end{flushleft}

\begin{enumerate}
\item The Penrose diagrams for (a) the generic BTZ black hole;
(b) the static ($J\!=\!0$) black hole; and (c) the extreme
($J\!=\!\pm M\ell$) black hole.
\item The phase diagram for a collapsing dust shell with initial
mass $\mu$, initial radius $r_0$, and initial velocity $\dot r_0$.
The end point in region I is open conical adS space; in region II it
is the BTZ exterior metric; in region III it is the BTZ interior
metric; and in region 4 it is closed conical adS space.
\item The upper half-space representation of the Euclidean black
hole.  The two hemispheres are to be identified along lines such
as $L$.  (The outer hemisphere has been cut open in this figure to
show the inner hemisphere.)
\item A region of the Euclidean black hole between $\tau\!=\!\tau_1$
and $\tau\!=\!\tau_2$.  The two nontrivial holonomies correspond to
the section of the line segment $K$ lying between the extreme values
of $\tau$, and the portion of the line segment $L$ lying between the
inner and outer hemispheres.  (The latter is a closed curve, since
the hemispheres are identified.)
\end{enumerate}

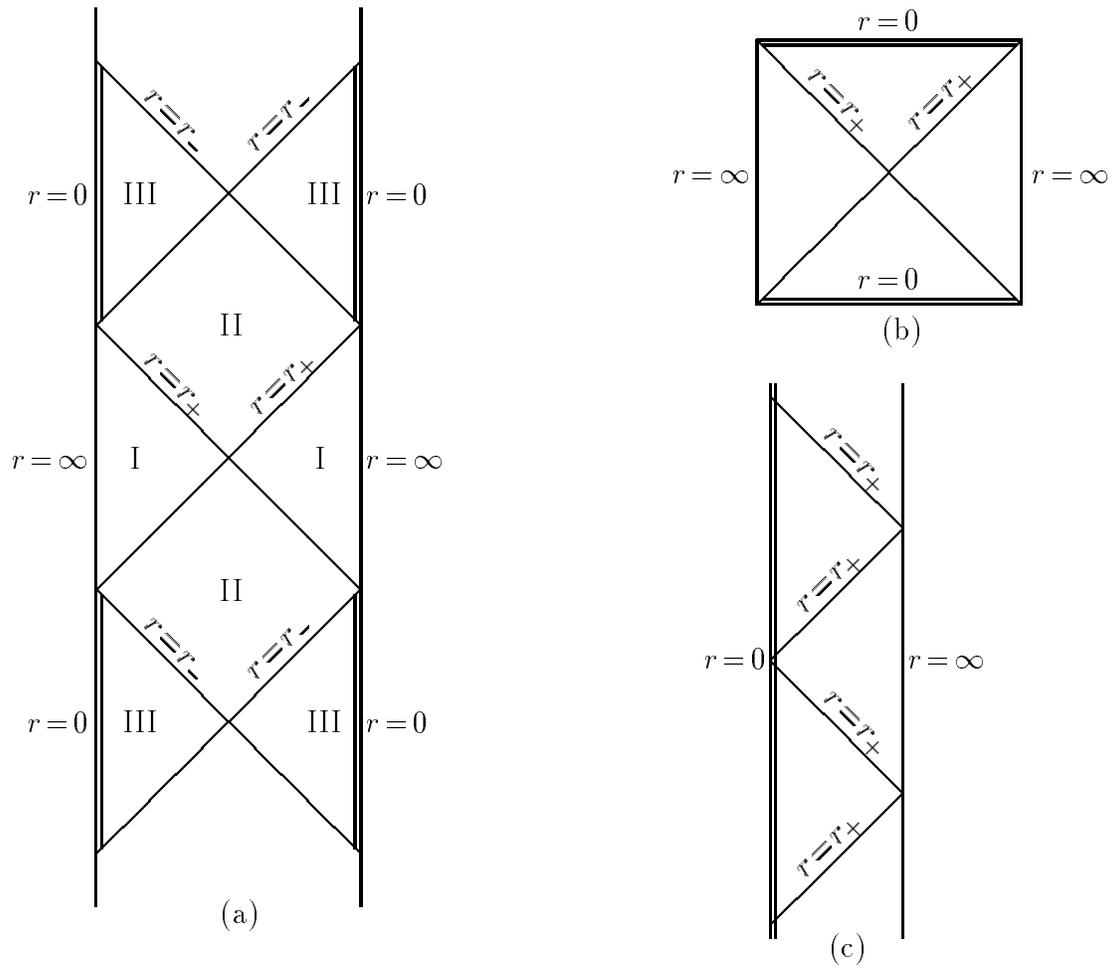
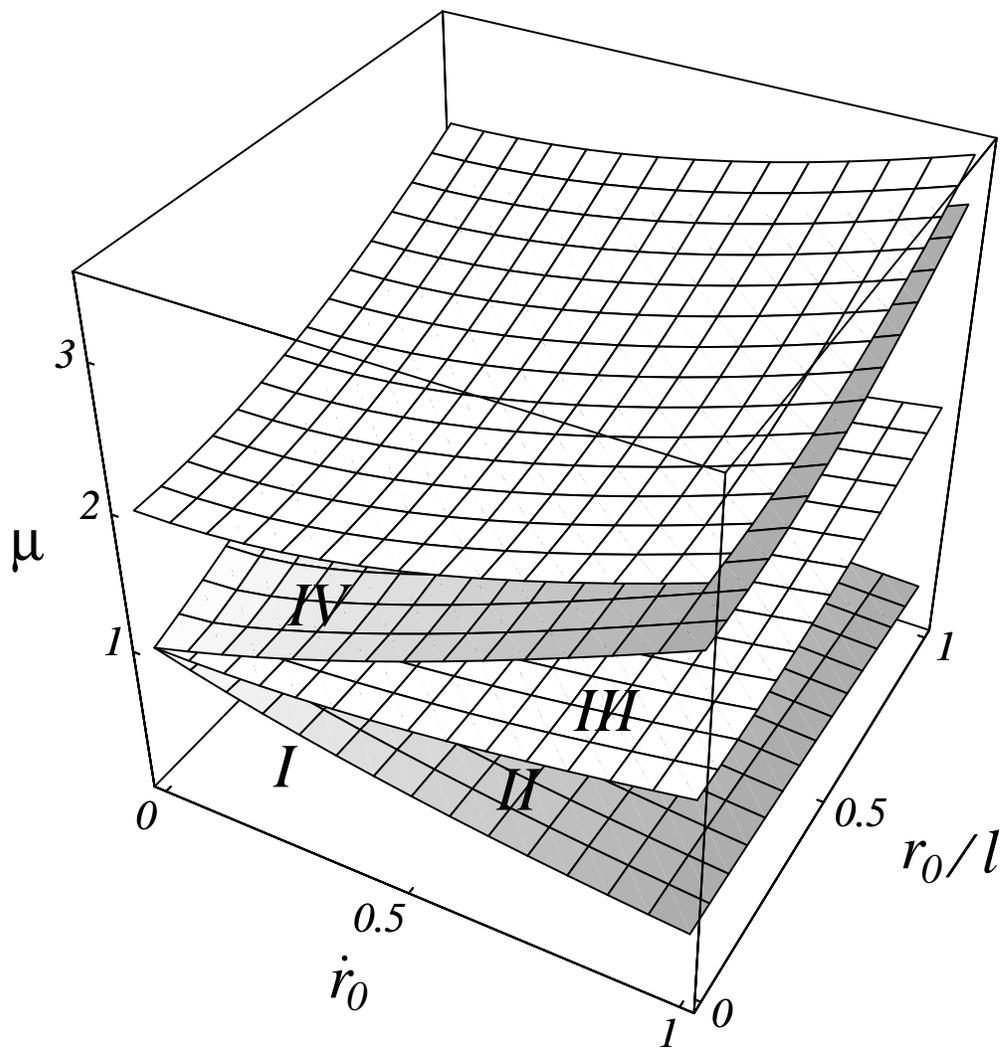
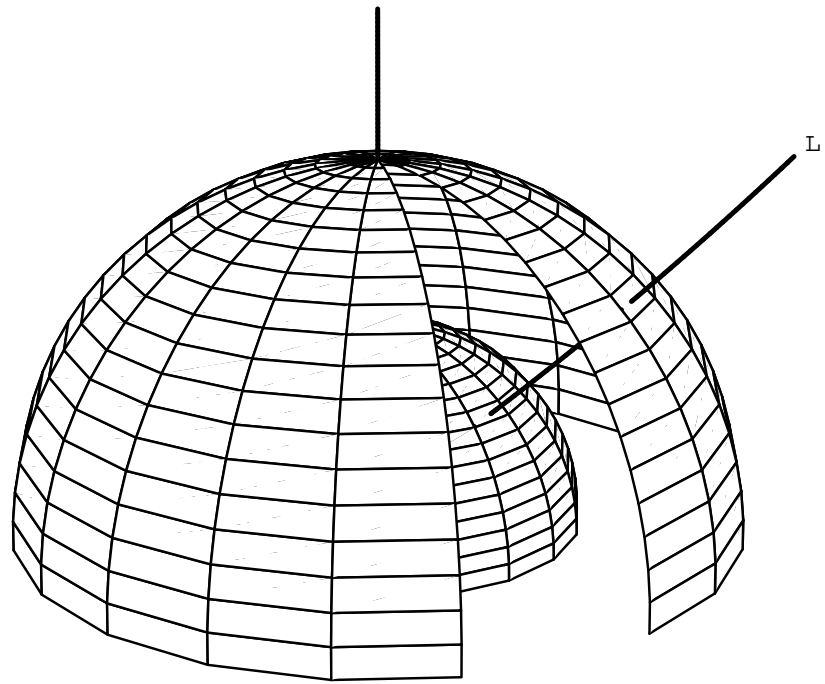
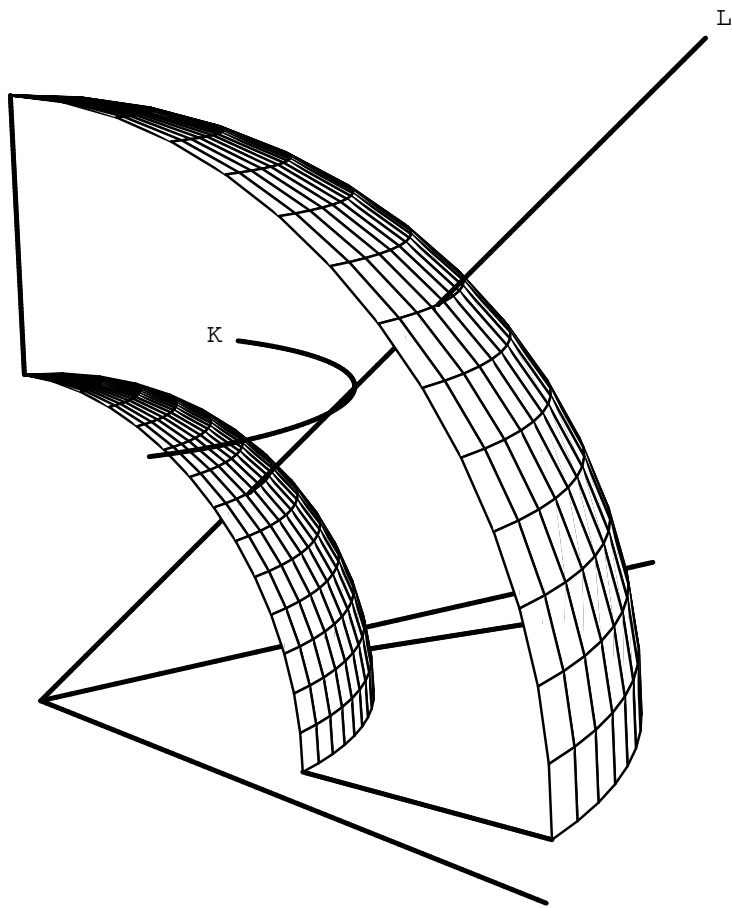
\end{document}